\documentclass[letterpaper,12pt]{book}
\pdfoutput=1
\newcommand{\usepackagewithoutwarning}[2][NoOptionsToBeGiven]{%
\let\OldPackageWarningNoLine\PackageWarningNoLine
\renewcommand{\PackageWarningNoLine}[2]{}
\ifthenelse{\equal{#1}{NoOptionsToBeGiven}}{\usepackage{#2}}{\usepackage[#1]{#2}}
\let\PackageWarningNoLine\OldPackageWarningNoLine
}

\usepackage[margin=1in,left=1.5in,head=14.5pt,includehead]{geometry}
\usepackage{amsfonts}
\usepackage{graphicx}
\usepackage[version=3]{mhchem}
\usepackage{tabularx}
\usepackage{fancyhdr}
\usepackage{emptypage}
\usepackage{ifthen}
\usepackage{setspace}
\usepackage{layout}
\usepackage{etoolbox}
\usepackage{booktabs}
\usepackage{cite}
\usepackage[super]{nth}
\usepackage[titletoc]{appendix}
\usepackage[flushmargin]{footmisc}
\usepackage[version=3]{mhchem}
\usepackagewithoutwarning{tocstyle}
\usepackage[colorlinks=true,linkcolor=black,citecolor=black, filecolor=black,urlcolor=black]{hyperref}

\makeatletter
\def\@part[#1]#2{%
    \ifnum \c@secnumdepth >-2\relax
      \refstepcounter{part}%
      \addcontentsline{toc}{part}{\partname~\thepart\hspace{1em}#1}%
    \else
      \addcontentsline{toc}{part}{#1}%
    \fi
    \markboth{}{}%
    {\centering
     \interlinepenalty \@M
     \normalfont
     \ifnum \c@secnumdepth >-2\relax
       \huge\bfseries \partname\nobreakspace\thepart
       \par
       \vskip 20\p@
     \fi
     \Huge \bfseries #2\par}%
    \@endpart}
\makeatother

\makeatletter
\def\unnumfootnote{\xdef\@thefnmark{}\@footnotetext}
\makeatother

\long\def\symbolfootnote[#1]#2{\begingroup%
\def\thefootnote{\fnsymbol{footnote}}\footnote[#1]{#2}\endgroup}

\usetocstyle{standard}
\settocfeature{raggedhook}{\raggedright}
\settocstylefeature[-1]{entryhook}{\Large\bfseries}

\let\stdthebibliography\thebibliography
\let\stdendthebibliography\endthebibliography
\renewenvironment*{thebibliography}[1]{%
  \stdthebibliography{99}}
  {\stdendthebibliography}

\newlength{\chaptervskip}
\makeatletter
\def\@makechapterhead#1{%
  \vspace*{\chaptervskip}%
  {\parindent \z@ \raggedright \normalfont
    \ifnum \c@secnumdepth >\m@ne
        \huge\bfseries \@chapapp\space \thechapter
        \par\nobreak
        \vskip 20\p@
    \fi
    \interlinepenalty\@M
    \Huge \bfseries #1\par\nobreak
    \vskip 40\p@
  }}

\def\@makeschapterhead#1{%
  \vspace*{\chaptervskip}%
  {\parindent \z@ \raggedright
    \normalfont
    \interlinepenalty\@M
    \Huge \bfseries  #1\par\nobreak
    \vskip 40\p@
  }}
\makeatother

\renewcommand{\chaptermark}[1]{ \markboth{#1}{} }
\renewcommand{\sectionmark}[1]{ \markright{: #1} }
\let\oldsection\section
\renewcommand{\section}[2][UseTheFullSectionTitle]{\ifthenelse{\equal{#1}{UseTheFullSectionTitle}}{\oldsection[#2]{#2 \sectionmark{#2}}\sectionmark{#2}}{\oldsection[\texorpdfstring{#2}{#1}]{#2 \sectionmark{#1}}\sectionmark{#1}}}

\fancypagestyle{frontplain}{%
\fancyhf{} 
\fancyfoot[C]{\thepage} 
\renewcommand{\headrulewidth}{0pt}
\renewcommand{\footrulewidth}{0pt}}

\newcommand{\longchapter}[2]{\chapter{\texorpdfstring{#1}{#2}}\chaptermark{#2}}

\newcommand{\leftsub}[2]{{\vphantom{#2}}_{#1}{#2}}		
\newcommand{\sfrac}[2]{\left.#1\middle/#2\right.}		
\newcommand{\br}[1]{\left(#1\right)}						
\newcommand{\fbr}[1]{\!\br{#1}}							
\newcommand{\rconf}{\mathfrak{R}}						
\newcommand{\overunder}[2]{\genfrac{}{}{0pt}{}{#1}{#2}}	
\newcommand{\abs}[1]{\left|#1\right|}					
\newcommand{\unit}[1]{\,\mathrm{#1}}						

\newcommand{\ket}[1]{\left| #1\right\rangle}

\newcommand{\unnumberedchapter}[1]{\cleardoublepage
\phantomsection
\addcontentsline{toc}{chapter}{#1}
\chapter*{#1}\chaptermark{#1}}

\let\oldsqrt\sqrt
\def\sqrt{\mathpalette\DHLhksqrt}
\def\DHLhksqrt#1#2{%
\setbox0=\hbox{$#1\oldsqrt{#2}$}\dimen0=\ht0
\advance\dimen0-0.2\ht0
\setbox2=\hbox{\vrule height\ht0 depth -\dimen0}%
{\box0\lower0.4pt\box2}}

\makeatletter
\def\@cite#1#2{{#1\if@tempswa , #2\fi}}
\makeatother
\let\citenum\cite
\renewcommand{\cite}[1]{[\citenum{#1}]}

\graphicspath{ {./figures/} }

\def\doctitle{Heavy Rydberg Photo-dissociation Cross-section Calculations and Experimental Progress Towards Cold Collisions in Lithium}
\def\docauthor{Lisa Madeleine Ugray}

\begin{document}
\title{\doctitle}
\author{\docauthor}
\frontmatter
\setlength{\chaptervskip}{-40pt} 
\begin{titlepage}
 \begin{center}
  \vspace*{0.5in}
  \begin{spacing}{1.3}
   {\LARGE\textsc{\doctitle}}
  \end{spacing}
  \vspace{4.5in}

  TRENT UNIVERSITY
  
  Peterborough, Ontario, Canada
  
  {\copyright} 2013 \docauthor , some rights reserved.
  
  This work is released under a \href{http://creativecommons.org/licenses/by-sa/3.0/}{Creative Commons Attribution Share Alike 3.0 License}
  
  \url{http://creativecommons.org/licenses/by-sa/3.0/}
  
  Materials Science M.Sc. Graduate Program
  
  January 2014
 \end{center}

\end{titlepage}
\setcounter{page}{2}
\unnumberedchapter{Abstract}
\begin{center}
 \textbf{\doctitle}\\
 
 by \docauthor
\end{center}

\noindent This thesis is divided into two parts, each of which supports constructing and using a lithium magneto-optical trap for cold collision studies:\\

\noindent\textbf{\autoref{part:rydberg}}

One outgoing channel of interest from cold collisions is the production of ion pairs.  We describe an effective method for calculating bound-to-continuum cross-sections for charged binary systems by examining transitions to states above the binding energy that become bound when the system is placed within an infinite spherical well.  This approach is verified for ionization of a hydrogen atom, and is then applied to the heavy Rydberg system \ce{Li+...I-}.\\

\noindent\textbf{\autoref{part:wavemeter}}

A wavemeter previously built in the lab is redesigned for increased reliability and ease of use by replacing the optical hardware with a rocker system, which can be aligned in mere minutes rather than half a day as was previously the case.  The new wavemeter has been tested through saturated absorption spectroscopy of lithium.\\


\unnumberedchapter{Acknowledgements}
The work presented in this thesis all took place in the Optical Physics group at Trent University under the supervision of Prof. Ralph Shiell.  It was through working with him that I first began to undertake scientific research as a co-op student in high school; I owe him a great debt of gratitude for the encouragement and guidance he has provided me over nine years, both before and during my masters, in the pursuit of scientific understanding.

Thanks are also due to the many people with whom I collaborated or who performed preliminary work from which the matter of this thesis developed:  Eric Brown and Jaclyn Semple did some calculations that lay the groundwork for \autoref{ch:rydberg} of this thesis, and Bill Atkinson provided critical reading of a manuscript from which the chapter was developed.  Julian Atfield, Tom McCarthy and Ralph Shiell were my coauthors on a paper describing a wavemeter built in the lab, the improvement of which I describe in \autoref{ch:wavemeter}. Ed Wilson in the Science Workshop provided the dial guage which I describe in the same chapter.

I must also take a moment to thank Jeffrey Philippson.  Although we did not work collaboratively on any of the work presented in this thesis, he was always helpfully willing to be a sounding board for ideas, and I enjoyed and profited from our discussions scientific and philosophical on the cycle ride back downtown at the end of a day of work at Trent.

Finally I wish to thank my parents, Cecily Ugray and David Roberts, without whose support this thesis would never have come to be.

\unnumberedchapter{Publications}
Much of the content of \autoref{ch:rydberg} of this thesis has been reprinted with permission from L.~M. Ugray and R.~C. Shiell, ``Elucidating Fermi's Golden Rule via bound-to-bound transitions in a confined hydrogen atom,'' Am. J. Phys. \textbf{81}, 206-210 (2013). Copyright 2013, American Association of Physics Teachers.

\autoref{ch:wavemeter} of this thesis extends work published previously by the author in M. Ugray, J.~E. Atfield et~al., ``Microcontroller-based wavemeter using compression locking of an internal mirror reference laser,'' Rev. Sci. Instr. \textbf{77}, 113109 (2006).

\makeatletter
\@openrightfalse
\makeatother

\tableofcontents

\clearpage
\phantomsection
\addcontentsline{toc}{chapter}{\listfigurename}
\listoffigures
\markboth{}{}

\makeatletter
\@openrighttrue
\makeatother
\newpage
\pagestyle{frontplain}
\ \cleardoublepage

\pagestyle{fancy}
\mainmatter
\fancypagestyle{plain}{%
\fancyhf{} 
\fancyhead[RO,LE]{\thepage} 
\renewcommand{\headrulewidth}{0pt}
\renewcommand{\footrulewidth}{0pt}}

\setlength{\chaptervskip}{50pt} 
\fancyfoot{}
\fancyhead[RO,LE]{\thepage}

\chapter{Introduction}\label{ch:intro}

The relevantly recent advent of quantum theory forced physicists to radically rethink the fundamental nature of the universe.  While Newtonian mechanics matches our intuitive and experiential understanding of how systems should behave, the same cannot be said of quantum mechanics which deals with systems too small to be a part of the daily human experience.

The divide between the macroscopic world of Newtonian mechanics and the microscopic world of quantum mechanics is a fascinating one.  The correspondence principle provides a bridge between the two theories, and those systems which push the boundaries, coherent quantum systems with sizes approaching those which would generally be described by classical physics, are of particular interest.

Bose-Einstein condensates of atomic gases \cite{einstein,anderson,anglin,wijngaarden}, in which a large population of atoms is rendered coherent by being forced through low temperature into the ground state, are one way in which this boundary is being pushed.  Weakly-bound ion-pairs, \ce{A+...B-} with their very large bond lengths can also be used to explore this boundary.  These systems of ion pairs which are bound primarily by Coulomb forces are called \emph{heavy} Rydberg systems \cite{reinhold} to distinguish them from traditional \emph{electronic} Rydberg systems \ce{A+...$e$-} composed of a bound nucleus and electron which are likewise bound primarily by Coulomb forces.  Given their large size yet simple binary structure, heavy Rydberg systems are prime candidates for studies which bridge quantum and classical physics.  Heavy Rydberg systems have been formed in molecular beam experiments \cite{suits}, in room temperature collisions \cite{kalamarides}, and have been proposed in cold collision environments \cite{kirrander}.  This latter approach, with minimal Doppler broadening, is that which is the main thrust of the present research group.

This thesis is divided into two main and distinct parts, which are unified by a common goal of working towards the construction of a magneto optical trap for lithium and its use in the study of cold collisions in lithium.

In the intended cold collision experiment, many final channels are energetically possible. Our primary interest is a search for weakly bound lithium ion pairs or heavy Rydberg systems (denoted \ce{Li+...Li-}), which we would like to study.  It is reasonable to assume that these may be strong absorbers of long wavelength room temperature black-body radiation as is the case for traditional highly-excited Rydberg systems \cite{cooke}.  To that end, it is of interest to determine the lifetime of the species that will be created which will dictate the time-scale on which any studies must be performed.  In order to lay the groundwork for these calculations, \autoref{part:rydberg} is a theoretical examination of the stabilities of binary charged systems against photo-transitions to the continuum.  Though more commonly called dissociation or ionization for the separation of nuclei or electron from core respectively, it can be convenient to refer to these events as \emph{charge separation} to reflect this similarity.  \autoref{part:rydberg} begins by examining photo-ionization of hydrogen in order to test our chosen numerical method and show that it generates a cross-sectional function which agrees with that found using an exact analytic method.  The numerical method is then used to perform the same calculation for \ce{Li+...I-}, a species for which the energies of some states have already been calculated \cite{pan}, allowing our methods to be verified at an intermediate step.  Convergence is again shown for increasing radius of confinement.  To the best of our knowledge, ours is the first heavy Rydberg photo-dissociation cross section calculations presented.

This lays excellent groundwork for minor adjustments to be made to the parameters of the program found in \autoref{app:code} in order to calculate photo-dissociation cross-sections for \ce{Li+...Li-}, or indeed any other ion pair.   While we would have ideally preferred to perform the calculation for \ce{Li+...Li-} directly, a similar intermediate calculation could not be found in the literature for this species, as this is the first such calculation being performed, intermediate checks were deemed valuable.  Additionally, the oscillatory nature of the \ce{Li+...I-} cross-section that results from the calculations in \autoref{ch:ionpair} suggests the possibility of selective control using light which cannot ionize the ion pair but only force a particular desired downward transition.

In order to be able to study cold collisions in lithium, it is necessary to tune the diode lasers to be used to the correct frequency, $\approx 446\,800\unit{GHz}$, for cooling lithium atoms.  Since these diode lasers can be tuned over hundreds of gigahertz \cite{philippson}, searching blindly for a transition with a width on the order of megahertz is not feasible.  A tool for measuring the wavelength or frequency of laser light, a wavemeter, was previously built in the lab to address this problem, but was a highly impractical device to use, requiring many hours of work to align properly due to insufficient linearity in the track of the travelling Michelson interferometer that it employed.  To tune an external cavity diode laser to the correct wavelength, gross adjustments of the external grating to produce large changes in output frequency are generally required.  These adjustments can have the unintended consequences of changing the alignment of the output beam, which in turn requires realigning the beam into the wavemeter.  Additionally, in practice multiple readily available laser diodes of promising nominal wavelengths must be purchased and their free-running (without external cavity) wavelength measured to determine which ones may be tuned to the desired wavelength.  Given the necessity of being able to quickly check or recheck the wavelength of the laser, a major redesign and construction of a new wavemeter was necessary to advance the experiment.  This was accomplished and is presented in \autoref{part:wavemeter} in which the optical layout and hardware of the initial wavemeter are completely replaced, and the signal processing electronics are improved to remove false signals produced by cross-talk in the previous system.  The new wavemeter is shown to be accurate through the use of saturated absorption spectroscopy \cite{demtroder}, applied to lithium.  Verifying the frequencies reported by the wavemeter through the use of saturated absorption spectroscopy also serves a secondary purpose, as it permits the next step in the experiment to be performed more readily: that of locking the laser to the saturated absorption spectrum at the frequency found using the wavemeter.

\part{Calculating the stabilities of Rydberg-like systems against bound-to-continuum photo-transitions}\label{part:rydberg}

\chapter{Literature Review: Photo-transitions and Rydberg systems}
The \emph{cross-section} \cite{bethe}, $\sigma\fbr{\omega}$, is commonly used to quantify the absorption of incident light of angular frequency $\omega$ by a sample comprising a set of particles (atoms, molecules, \emph{etc.}). It is called a cross-section since it is an effective cross-sectional area that each particle presents on average to an incoming flux of photons, such that if a photon considered to be a point-like particle were to pass through that effective area, it would be absorbed.  Another common measure of the rate at which particles absorb light, the oscillator strength distribution \cite{gallagher}, $\frac{df}{dW}$, where W is the internal energy of the system, is linearly related to the cross-section by
\begin{equation}
 \frac{df}{dW}=\frac{c\hbar}{8\mathrm{a}_0^3\mathrm{Ry}^2\pi^2}\sigma .
\end{equation}

Calculating cross-sections for transitions within binary charged systems to the continuum requires choosing an appropriate normalization method of the continuum state involved.  One such normalization is an \emph{energy normalization} \cite{landau} which may on occasion generate exact analytical results.  A second method involves placing the system in an infinite potential well, \emph{box normalization} \cite{tellinghuisen}, such that states at all energies are bound, with states in the previously continuous regime becoming discretized due to the box.  Any results found using this method will converge towards the correct value for an increasing radius of confinement; the case as the radius of confinement approaches infinity is that of the unbound system.

The model of the confined hydrogen atom was initially developed in 1937 \cite{michels} to study the effects of high pressure on matter.  Since then it has been further studied to determine the energy levels of the confined atom: Wilcox \cite{wilcox} uses an approximate formula to show breaking of degeneracy caused by confinement in which states with the lowest angular momentum raise most in energy.  Laughlin \emph{et al.} \cite{laughlin} made creative use of perturbation theory for radii of confinement close to nodes of wavefunctions for the unconfined atom.  Aquino \emph{et al.} \cite{aquino} showed that early \nth{21} century computers are up to the task of calculating the energies of the states of the confined hydrogen atom up to 100 significant figures, and also reported its polarizability for a few radii of confinement. Mazie \emph{et al.} \cite{maize} also examined the polarizability of the atom as did Laughlin \cite{laughlin2004}, who also examined the 1s$\to$2p oscillator strength.  Stevanovi\'c \cite{stevanovic} too examined the oscillator strength, extending Laughlin's work to bound-to-bound transitions for $n\le 5$.  Dolmatov \cite{dolmatov} examined photoionization of confined atoms, but focused on real world confinement structures such as \ce{C60}, and was interested in the confined atom in such structures for its own sake.  Conversely, in what follows we use confinement as a mathematical tool to discretize continuum states, and then allow the radius of confinement to increase, tending towards infinity, to understand the nature of the unconfined system.

\chapter{Cross-sections for photo-ionization of hydrogen}\label{ch:rydberg}

\section{Introduction}

The\label{sec:intro}\symbolfootnote[0]{Much of the content of this chapter has been reprinted with permission from L.~M. Ugray and R.~C. Shiell, ``Elucidating Fermi's Golden Rule via bound-to-bound transitions in a confined hydrogen atom,'' Am. J. Phys. \textbf{81}, 206-210 (2013). Copyright 2013, American Association of Physics Teachers.  This copyright notice must appear in any derivative or reproduced materials.} intended cold collision experiment will study collisions of lithium atoms, chosen for their relatively simple structure, and the existence of a convenient cooling transition.  A collision of interest is that between a ground state lithium atom, and an excited lithium atom, which has many energetically possible final channels, such as excited molecular or ionic \ce{Li_2}. Our primary interest is a search for \emph{weakly} bound lithium ion pairs, \ce{Li+...Li-}.  As these ion pairs will have some of the characteristics of Rydberg atoms, it is quite possible that they will be strong absorbers of long wavelength light from background blackbody radiation.  It will be important to understand the (undesirable) photo-ionization processes that may occur and their rates.  To that end, we examined Fermi's golden rule \cite{sakurai}:
\begin{equation}\label{eq:fermirule}
 \Gamma_{i\rightarrow \{j\}}=\frac{2\pi}{\hbar}\overline{\left|\left<j\right|\hat{H}'\left|i\right>\right|^2}\rho\fbr{W},
\end{equation}
This gives the transition rate from an initial bound state $i$ to the infinitude of continuum states $\{j\}$ in a small energy range centred about $W$ with $\rho\fbr{W}$ being the density of states in the energy range with the same symmetry as $j$.  While this is often presented \cite{fermi} for a time-independent perturbation, $\hat{H}'$, this rule can also apply for a sinusoidal perturbation, $\mathcal{\hat{H}}'\fbr{t}$, with angular frequency $\omega$, by setting
\begin{equation}\label{eq:fermitimedep}
 \hat{H}'=\frac{\mathcal{\hat{H}}'\fbr{t}}{e^{i\omega t}+e^{-i\omega t}},
\end{equation}
rendering $\hat{H}'$ again time-independent \cite{atkins,starace}\footnote{Note that for more general electromagnetic wave packets, a full-time-dependent analysis is required; see Refs.~\citenum{seideman} and \citenum{yeazell} for more details.}. However, when applying Fermi's Golden Rule it is not immediately clear how to interpret the product of $\rho\fbr{W}$ and the mean of the square of the matrix element, given that the former is infinite for states in the continuum where all energies are allowed, and the latter incorporates the space-normalized wavefunctions $\{\left<j\right|\}$, which have zero amplitude as they extend over all space.

To resolve this conundrum and understand how best to use Fermi's Golden rule to determine the time-scale on which the experiment must be performed before population is lost, this chapter examines the cross-section for photo-absorptions in a few situations.  The cross-section is used since this is a commonly reported quantity that has consistent meaning for both bound-to-bound transitions, and the bound-to-continuum transitions such as the photo-dissociation of ion pairs that are ultimately the processes of interest. I start in \autoref{sec:btb} by examining the cross-section, $\sigma\fbr{\omega}$, for a bound-to-bound transition using the hydrogen atom as an example since its wavefunctions are well-known.  In \autoref{sec:btc} one method for calculating bound-to-continuum cross-sections using energy-normalized wavefunctions is examined, maintaining the example of hydrogen for which analytical energy normalized continuum wavefunctions are known.  In \autoref{sec:wellH} a second method of calculating bound-to-continuum cross-sections is examined, again using the example of hydrogen so that the results can be compared with those from \autoref{sec:wellH} and be shown to agree.

Dealing with atomic hydrogen has made it convenient to use two commonly encountered constants throughout this chapter and the following: the Bohr radius $\mathrm{a}_0=\sfrac{4\pi\epsilon_0\hbar^2}{\br{m_ee^2}}\allowbreak\approx5.29{\times}10^{-11}\unit{m}$, and the Rydberg unit of energy $\mathrm{Ry}=m_ee^4/[\br{4\pi\epsilon_0}^22\hbar^2] \allowbreak\approx2.18{\times}10^{-18}\unit{J}$.

\section[Bound-to-bound transitions]{Cross-sections for bound-to-bound transitions}\label{sec:btb}

A physical parameter that quantifies the absorption of light of frequency $\omega$ by a sample of atoms with some distribution of initial states is the \emph{cross-section}, $\sigma\fbr{\omega}$.  The average photon absorption rate per atom due to an incident narrowband photon flux, $F$ (photons/unit time/unit area), is given by $\Gamma=\sigma\fbr{\omega}F$ \cite{bethe}, and thus the cross-section indicates the average effective area of an atom presented to the light at angular frequency $\omega$.  The number of absorptions from a single weak (i.e. assuming negligible population inversion) incident beam of cross-sectional area A travelling in the positive $x$-direction within a thickness $dx$ of a sample with number density of all particles $N$ in a time $dt$ is $\br{NAdx}\Gamma dt=NAdx\sigma Fdt$.  This is equal to the decrease in the number of photons present: $-dFdtA$.  Cancelling yields $N\sigma Fdx=-dF$ which can be rearranged as $\sfrac{dF}{dx}=-\sigma FN$.  Since the intensity of the light, $I$, differs from $F$ by only a constant factor ($\hbar\omega$), this yields another convenient expression defining the cross-section,
\begin{equation}\label{eq:sigmadef}
 \frac{dI}{dx}=-\sigma\fbr{\omega} NI.
\end{equation}

For bound-to-bound transitions between two particular quantum states, a standard semi-classical treatment using the time-dependent Schr\"odinger equation for a narrowband, weak incident light beam gives \cite{bernath}
\begin{equation}
 \Gamma_{i\rightarrow j}=\frac{\pi\omega}{c\epsilon_0 \hbar}\left|\left<j\right|\boldsymbol{\hat{d}}\left|i\right>\right|^2g\fbr{\omega-\omega_0}F_\epsilon,
\end{equation}
where $\boldsymbol{\hat{d}}$ is the electric dipole moment operator due to all charges in the atom, $F_\epsilon$ is the incident photon flux with polarization $\epsilon$ which couples the transition $i{\rightarrow}j$, and $g\fbr{\omega-\omega_0}$ is the lineshape of the transition with $\int\!g\,d\omega=1$.  In general, however, many atomic levels are degenerate and light can couple many of these states, so the average absorption rate per atom for an ensemble of atoms is given by
\begin{equation}
\Gamma=\sum\limits_{\{i,\,j\}}\Gamma_{i\rightarrow j}\sfrac{N_i}{N},
\end{equation}
where the summation runs over all pairs of states coupled by the incident light, and $\sfrac{N_i}{N}$ is the fraction of atoms in each initial state $\left|i\right>$.

It is usual to report absorption cross-sections that assume the atoms are evenly distributed among all $2J+1$ possible $M$ initial states.  This distribution is frequently encountered, and $\Gamma$ is then by symmetry independent of the polarization characteristics of the incident light.  For ease in deriving this cross-section, we will arbitrarily consider incident light which is linearly polarized along the $z$-axis ($\pi$-polarized, so $\Delta M{=}0$).  For a transition $\gamma J{\rightarrow}\gamma'J'$, where $\gamma$ represents all quantum numbers other than $J$ and $M$, this gives
\begin{equation}
\sigma\fbr{\omega}=\frac{\pi\omega}{c\epsilon_0 \hbar}\frac{g\fbr{\omega-\omega_0}}{2J+1}\sum\limits_{M}\left|\left<\gamma'J'M\right|\hat{d}_z\left|\gamma JM\right>\right|^2,
\end{equation}
where $\hat{d}_z$ is the $z$-component of the electric dipole moment operator.  For the particular case of transitions $n\ell{\rightarrow}n'\ell'$ within the hydrogen atom (where $\gamma=n$, $J=\ell$, and $M=m$), neglecting electron and nuclear spin, this gives
\begin{align}
 \begin{split}\label{eq:hydrogensigma}
  \sigma\fbr{\omega}&=\frac{\pi\omega e^2}{c\epsilon_0 \hbar}\frac{g\fbr{\omega-\omega_0}}{2\ell+1}\left|D_{n\ell\rightarrow n'\ell'}\right|^2\sum\limits_{m}\left|\int\limits_0^{2\pi}\!\int\limits_0^\pi {Y_{\ell'}^{m}}^* \br{\sqrt{\frac{4\pi}{3}}Y_1^0} Y_\ell^m\mathrm{sin}\theta\,d\theta\, d\phi\right|^2\\
  &=\frac{\pi\omega e^2}{c\epsilon_0 \hbar}\frac{g\fbr{\omega-\omega_0}}{2\ell+1}\left|D_{n\ell\rightarrow n'\ell'}\right|^2\sum\limits_{m}\frac{\ell_\mathrm{max}^2-m^2}{4\ell_\mathrm{max}^2-1}\\
  &=\frac{\pi\omega e^2}{3c\epsilon_0 \hbar}\frac{\ell_\mathrm{max}}{2\ell+1}\left|D_{n\ell\rightarrow n'\ell'}\right|^2g\fbr{\omega-\omega_0},
 \end{split}
\end{align}
where the integral incorporating the standard spherical harmonics, $Y_\ell^m$, was evaluated using Ref.~\citenum{zare}, $\ell_\mathrm{max}$ is the greater of $\ell$ and $\ell'$, $\abs{\ell-\ell'}=1$, and the \emph{radial matrix element} is defined in terms of the well-known space-normalized radial wavefunctions, $R_{n\ell}$, by
\begin{equation}\label{eq:Dnlnl}
 D_{n\ell\rightarrow n'\ell'}=\int\limits_0^{\infty}R_{n'\ell'}^*r\,R_{n\ell}\,r^2\,dr.
\end{equation}

The \emph{integrated cross-section} is a commonly presented quantity that reflects the overall strength of a transition by integrating $\sigma$ over all incident frequencies, $\nu$, rather than the angular frequencies, $\omega=2\pi\nu$, used above.  It follows that this integrated cross-section is given by:

\begin{equation}
 \int\!\!\sigma\,d\nu=\frac{\omega_0 e^2}{6c\epsilon_0 \hbar}\frac{\ell_\mathrm{max}}{2\ell+1}\left|D_{n\ell\rightarrow n'\ell'}\right|^2.
\end{equation}
As an example, the integrated cross-section for the 1s$\to$2p transition of hydrogen is $\sfrac{32768\pi\unit{a_0}^3\unit{Ry}^2}{19683 \hbar^2 c}$.

\section[Energy-normalization approach]{Cross-sections for bound-to-continuum transitions: Energy-normalization}\label{sec:btc}

One approach to evaluating the problematic product described in \autoref{sec:intro} is to use energy-normalized wavefunctions for the continuum states, defined by \cite{landau}:
\begin{equation}\label{eq:energynormdef1}
 \int\limits_\mathrm{\overunder{all}{space}}\!\!\!\Psi_W^*\Psi_{W'}\, d\tau=\delta\fbr{W-W'},
\end{equation}
where $W$ is the energy of the continuum state, with $W=0$ corresponding to the energy being referenced to the ionization limit of the system.  Integrating over $W'$ on both sides for an arbitrarily small energy range, $\Delta W$, centred about $W$ gives another representation of this equation:
\begin{equation}\label{eq:energynormdef}
 \int\limits_\mathrm{\overunder{all}{space}}\!\!\!\Psi_W^*\!\!\!\!\int\limits_{W-\Delta W}^{W+\Delta W}\!\!\!\!\Psi_{W'} dW' d\tau=1.
\end{equation}
We first examine a hydrogen atom with \emph{space-normalized} continuum wavefunctions $j=Y_{\ell}^{m}R_{W\ell}$ (i.e. a radial wavefunction that necessarily has zero amplitude as it extends over all space) and show that multiplication by $\sqrt{\rho\fbr{W}}$ (which is infinite) gives \emph{energy-normalized finite-amplitude} wavefunctions satisfying  \autoref{eq:energynormdef}.  The left hand side of \autoref{eq:energynormdef} with such a substitution then takes the form
\begin{align}\label{eq:energynormproof}
 \begin{split}
  &\int\!\!\!\!\!\!\int\limits_\mathrm{\overunder{all}{space}}\!\!\!\!\!\!\int\!\! \sqrt{\rho}\,{Y_{\ell}^{m}}^*R_{W\ell}^*\!\!\!\!\!\!\!\int\limits_{W-\Delta W}^{W+\Delta W}\!\!\!\!\!\!\!\! \sqrt{\rho}\,Y_{\ell}^{m}R_{W'\ell}\,dW'\,d\tau\\
  =&\int\limits_0^\infty r^2R_{W\ell}^*\!\!\!\int\limits_{W-\Delta W}^{W+\Delta W}\!\!\! R_{W'\ell}\frac{dn}{dW'}\,dW'\,dr\\
  =&\int\limits_0^\infty r^2R_{W\ell}^*\!\int\limits_{n^-}^{n^+}\! R_{W'\ell}\,dn\,dr=\int\limits_0^\infty r^2R_{W\ell}^*\!R_{W\ell}\,dr\equiv 1,
 \end{split}
\end{align}
where we have used in the first step that $\rho$ is $\sfrac{dn}{dW}$ evaluated twice in \autoref{eq:energynormproof}, once at $W$ and once at $W'$, with $W'$ bounded by $W\pm\Delta W$, and in the final step the integral over $n$ is unity since there is only one (space-normalized) radial wavefunction $R_{W'\ell}$ which is not orthogonal to $R_{W\ell}$.

\begin{figure}[tb]
 \includegraphics[width=\columnwidth]{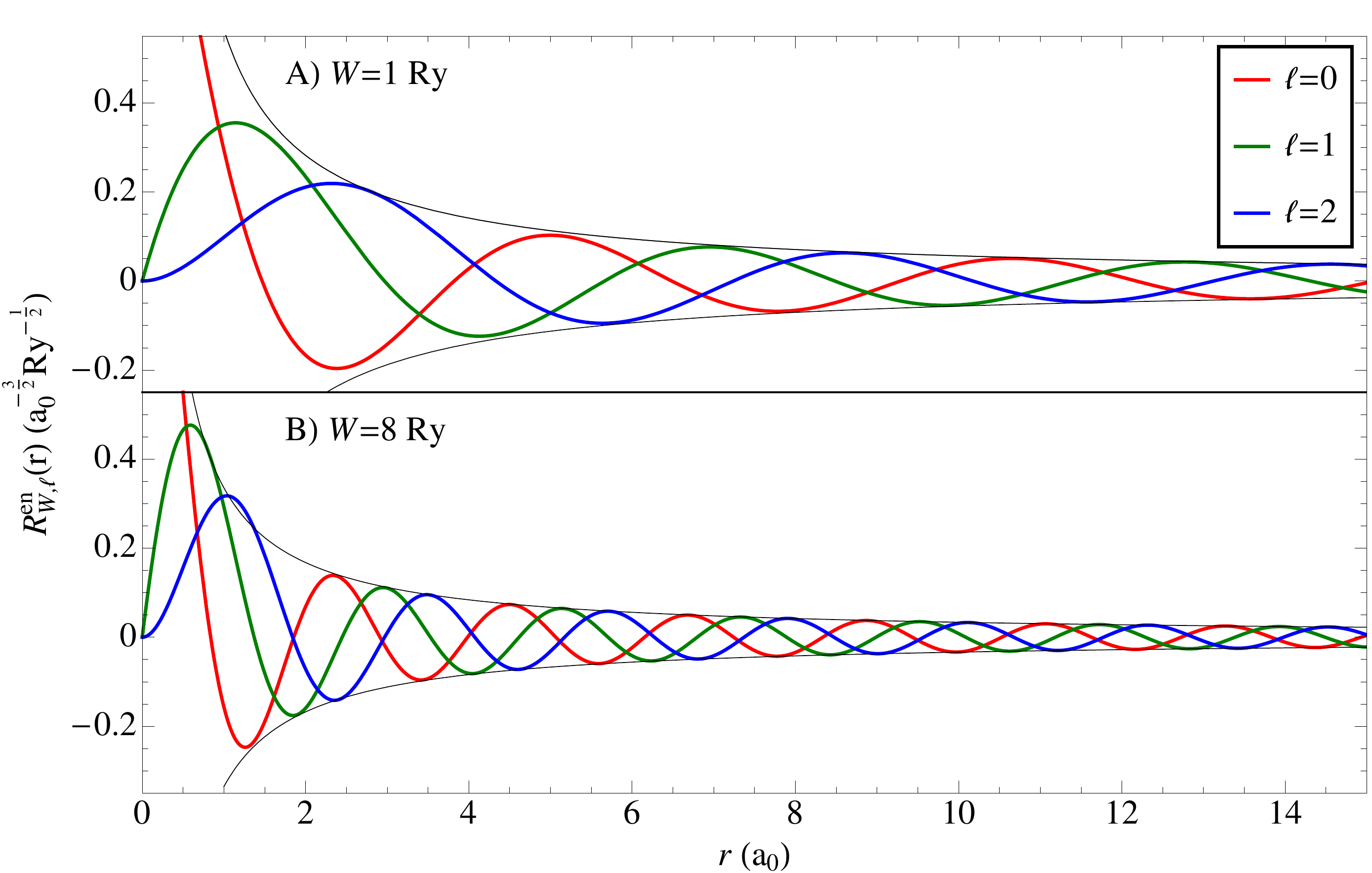}
 \caption[Plots of $R_{W\ell}^\mathrm{\,en}\fbr{r}$]{Plots of $R_{W\ell}^\mathrm{\,en}\fbr{r}$ for $\ell=0$ in red, $\ell=1$ in green, and $\ell=2$ in blue, at $W=1\unit{Ry}$ in plot A and $W=8\unit{Ry}$ in plot B.  The $W\mathrm{s}$ states have a finite value at $r=0$ of $\sqrt{\frac{2}{1-e^{\frac{-2 \pi}{\sqrt{\sfrac{W}{\mathrm{Ry}}}}}}}\unit{a}_0^{-\frac{3}{2}}\unit{Ry}^{-\frac{1}{2}}$. Both plots show an envelope $\sfrac{\pm\br{\mathrm{a}_0^2\pi^2W\mathrm{Ry}}^{\sfrac{-1}{4}}}{r}$ in black.}\label{fig:rplot}
\end{figure}

We therefore conclude that the photoionization cross-section for hydrogen can be derived from Fermi's Golden Rule by including the density of states within the bra: $\left<\sqrt{\rho}\,j\right|$, and using for this bra the energy-normalized continuum wavefunctions (energy \mbox{$W>0$}) with radial component \cite{burgess}:
\begin{align}\label{eq:normhydrogenabove}
 \begin{split}
  R_{W\ell}^\mathrm{\,en}\fbr{r}=&\sqrt{\frac{2\prod\limits_{s=0}^\ell \br{1+\frac{W}{\mathrm{Ry}}s^2}}{\br{1{-}e^{-\pi\sqrt{\frac{4\mathrm{Ry}}{W}}}}\mathrm{Ry}\,\mathrm{a}_0^3}}\,\frac{\br{\sfrac{2r}{\mathrm{a}_0}}^\ell}{\br{2\ell+1}!}e^{\frac{ir}{\mathrm{a}_0}\sqrt{\frac{W}{\mathrm{Ry}}}}\\
  &\quad\times\leftsub{1}{F}_1\fbr{\ell{+}1{-}i\sqrt{\frac{\mathrm{Ry}}{W}};\,2\ell{+}2;\,-2i\frac{r}{\mathrm{a}_0}\sqrt{\frac{W}{\mathrm{Ry}}}}.
 \end{split}
\end{align}
The normalization constants herein can be derived by applying \autoref{eq:energynormdef} to the asymptotic form of the radial wavefunction, and these result in wavefunctions that when multiplied by $r$ tend to an oscillatory function with constant amplitude of $\br{\mathrm{a}_0^2\pi^2W\mathrm{Ry}}^{\sfrac{-1}{4}}$ for large $r$ \cite{tellinghuisen}.  This can be seen in \autoref{fig:rplot}
where $R_{W\ell}^\mathrm{\,en}\fbr{r}$ is plotted for $\ell\in\left\{0,1,2\right\}$ at $W=1\unit{Ry}$ and $W=8\unit{Ry}$, together with an envelope showing $\pm\br{\mathrm{a}_0^2\pi^2W\mathrm{Ry}}^{\sfrac{-1}{4}}/r$ for each plot.  Note that when $W=-\sfrac{\mathrm{Ry}}{n^2}$, the functional form of \autoref{eq:normhydrogenabove}, which contains \mbox{$\leftsub{1}{F}_1\br{a;\,b;\,z}$}, the confluent hypergeometric function \cite{gradshteyn}, reduces to that of the well-known generalized Laguerre polynomials for hydrogenic bound states.

Again assuming a uniform distribution of initial $m$ states and employing $\pi$-polarized light with electric field amplitude $E_0$, and using the sinusoidal approach described in \autoref{eq:fermitimedep} to identify $\hat{H}'$ as $\sfrac{d_z E_0}{2}$, \autoref{eq:fermirule} becomes
\begin{align}
 \begin{split}
  &\Gamma_{n\ell\rightarrow W\ell}=\frac{1}{2\ell+1}\frac{2\pi}{\hbar}
  \sum\limits_m \left|\left<Y_{\ell'}^mR_{W\ell'}^\mathrm{\,en}\left|\frac{d_z E_0}{2}\right|Y_{\ell}^mR_{n\ell}\right>\right|^2.
 \end{split}
\end{align}  
By equating $E_0^2$ to $\sfrac{2\omega F_\pi\hbar}{c\epsilon_0}$, the photoionization cross-section can be seen to be:
\begin{align}
\begin{split}\label{eq:sigmacont}
  \sigma\fbr{\omega}=&\frac{4\pi^2\omega \mathrm{a}_0}{c}\frac{2\mathrm{Ry}}{2\ell+1}\left|D_{n\ell\rightarrow W\ell'}\right|^2\sum\limits_{m}\!\left|\int\limits_0^{2\pi}\!\int\limits_0^\pi\!{Y_{\ell'}^{m}}^*\!\br{\!\sqrt{\frac{4\pi}{3}}\!Y_1^0\!}\!Y_\ell^m\mathrm{sin}\theta\,d\theta\, d\phi\right|^2\\
  =&\frac{4\pi^2\omega \mathrm{a}_0 2\mathrm{Ry}}{3c}\frac{\ell_\mathrm{max}}{2\ell+1}\left|D_{n\ell\rightarrow W\ell'}\right|^2,
 \end{split}
\end{align}
where \mbox{$\hbar\omega=W+\sfrac{\mathrm{Ry}}{n^2}$.}  By analogy to \autoref{eq:Dnlnl}, we define
\begin{equation}\label{eq:DnlWl}
 D_{n\ell\rightarrow W\ell'}=\int\limits_0^{\infty}{R_{W\ell'}^\mathrm{\,en}}^{\hspace{-0.5em}*}\,r\,R_{n\ell}\,r^2\,dr,
\end{equation} 
which can be evaluated exactly for any $n$, $\ell$, and $\ell'$ to determine the photoionization cross-section as a function of energy.  Note that the units of \autoref{eq:DnlWl} are different from those of \autoref{eq:Dnlnl}, but the units of their pre-factors in Equations \ref{eq:sigmacont} and \ref{eq:hydrogensigma} respectively are also different, resulting in both cases with $\sigma\fbr{\omega}$ having units of area as expected.  We have calculated this photoionization cross-section from a variety of initial states in hydrogen and present them in the following section, demonstrating agreement with the result from box normalization calculations described below.

\section[Box normalization approach]{Cross-sections for bound-to-continuum transitions: Box normalization}\label{sec:wellH}
A second approach commonly proposed to resolve the mathematical dilemma described in \autoref{sec:intro} is to place the system in an infinite potential well of finite size, so that the problematic terms are neither zero nor infinite \cite{orear}.  We now demonstrate this approach, showing that the results converge to the correct values as the volume of the box tends to infinity.

The potential energy of a hydrogen atom contained in a spherical well of radius $r_0$ with origin located at the atom's center of mass is 
\begin{equation}
 U\fbr{r}=
\begin{cases}
 -\frac{2\unit{Ry}\unit{a}_0}{r} & r < r_0 \\
 \infty & r \ge r_0.
\end{cases}
\end{equation}

We start by solving the Schr\"odinger equation to find the allowed energies $W_{n\ell}\fbr{r_0}$ and energy eigenfunctions for such a confined hydrogen atom.  As the system remains spherically symmetric, the wavefunctions retain the standard $Y_\ell^m$ angular portion of the free hydrogen atom.  

\begin{figure}[tb]
 \includegraphics[width=\columnwidth]{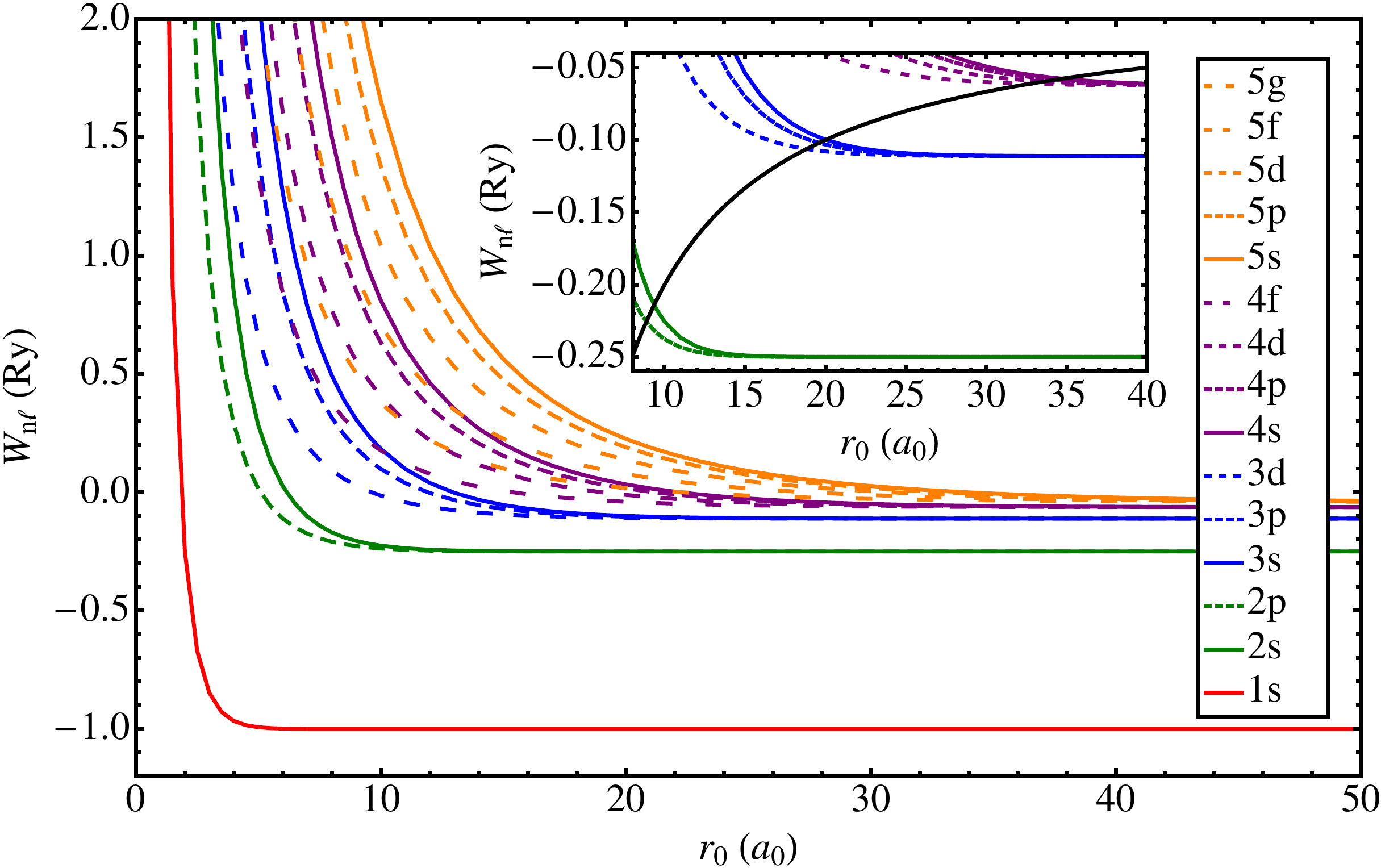}
 \caption[Energies of states as a function of radius of confinement]{The energies of states as a function of radius of confinement.  The inset shows an additional curve representing the potential energy as a function  of $r$ on the same scale.}\label{fig:energies}
\end{figure}

We adopt the functional form of \autoref{eq:normhydrogenabove} for each confined radial wavefunction, $\rconf_{W_{n\ell}\ell}$, and introduce the boundary condition:
\begin{equation}\label{eq:boundarycondition}
 \rconf_{W_{n\ell}\ell}\fbr{r_{\scriptscriptstyle 0}}\equiv 0.
\end{equation}
Allowed energies under this condition were found numerically for fixed $\ell$ and $r_0$ using Mathematica, and agree with those found previously by Aquino et al. \cite{aquino}. The boundary condition represented by \autoref{eq:boundarycondition} has the effect of raising the energies from those of the free atom bound states, as shown in \autoref{fig:energies}.
The inset illustrates that the energies begin to rise more dramatically when the radius of confinement, $r_0$, is such that the state no longer has a region of exponential decay.  States previously in the continuum no longer exist since properly speaking there is no continuum; instead an infinitude of bound states rise to energies greater than the ionization limit, leaving a finite number of (Coulombically) bound states below the ionization limit.  We call those states which rise in energy above the ionization limit \emph{pseudocontinuum} states, and relabel their energies $W$, dropping the subscript $n\ell$ to distinguish them from those of the (Coulombically) bound states.  It is important to note that in the region allowed by the box, the pseudocontinuum states have the exact same functional form as the true continuum states of the free atom at the same energy.

These bound and pseudocontinuum wavefunctions were then numerically space-normalized in the usual way.  We wish to find prefactors for each of the initial and final states which will have the property of causing the radial matrix element to tend, as $r_0{\rightarrow}\infty$, to that given in \autoref{eq:DnlWl} for the free atom.

We define the density of final states with a particular symmetry $\rho\br{W}$ for the confined atom to be $\sfrac{2}{\br{W^+{-}W^-}}$ where $W^+$ ($W^-$) is the energy of the adjacent state with the same symmetry above (below) $\left|\rconf_{W\ell}\right>$.  Thus, as $r_0$ goes to infinity so too will the density of states above the ionization limit.  Furthermore, from \autoref{sec:btc} as $r_0\rightarrow\infty$, pseudocontinuum states $\rconf_{W\ell}\br{r}$ assume zero amplitude in such a way that
\begin{equation}
 \sqrt{\rho\br{W}}\rconf_{W\ell}\fbr{r}\rightarrow R_{W\ell}^\mathrm{\,en}\fbr{r},\\
\end{equation}
and for (Coulombically) bound states, trivially
\begin{equation}
 \rconf_{W_{n\ell}\ell}\fbr{r}\rightarrow R_{n\ell}\fbr{r},
\end{equation}
where in each case $\rconf$, the confined-atom wavefunction, is space-normalized, and $R_{W\ell}^\mathrm{\,en}$ and $R_{n\ell}$ are energy-normalized and space-normalized, respectively.  We call $\sqrt{\rho\br{W}}\rconf_{W\ell}\fbr{r}$ a \emph{box normalized} wavefunction.  The free atom bound-to-continuum radial matrix element in \autoref{eq:DnlWl} can therefore be readily found from
\begin{align}\label{eq:curlyD}
 \mathfrak{D}_{W_{n\ell}\ell\rightarrow W\ell'} &\equiv \int\limits_0^{r_0}\sqrt{\rho\br{W}}\rconf_{W\ell'}^*\, r\mathfrak{R}_{W_{n\ell}\ell}\,r^2\,dr\\
 &\xrightarrow{r_0\rightarrow\infty} D_{n\ell\rightarrow W\ell'}.
\end{align}

\begin{figure}[t]
 \includegraphics[width=\columnwidth]{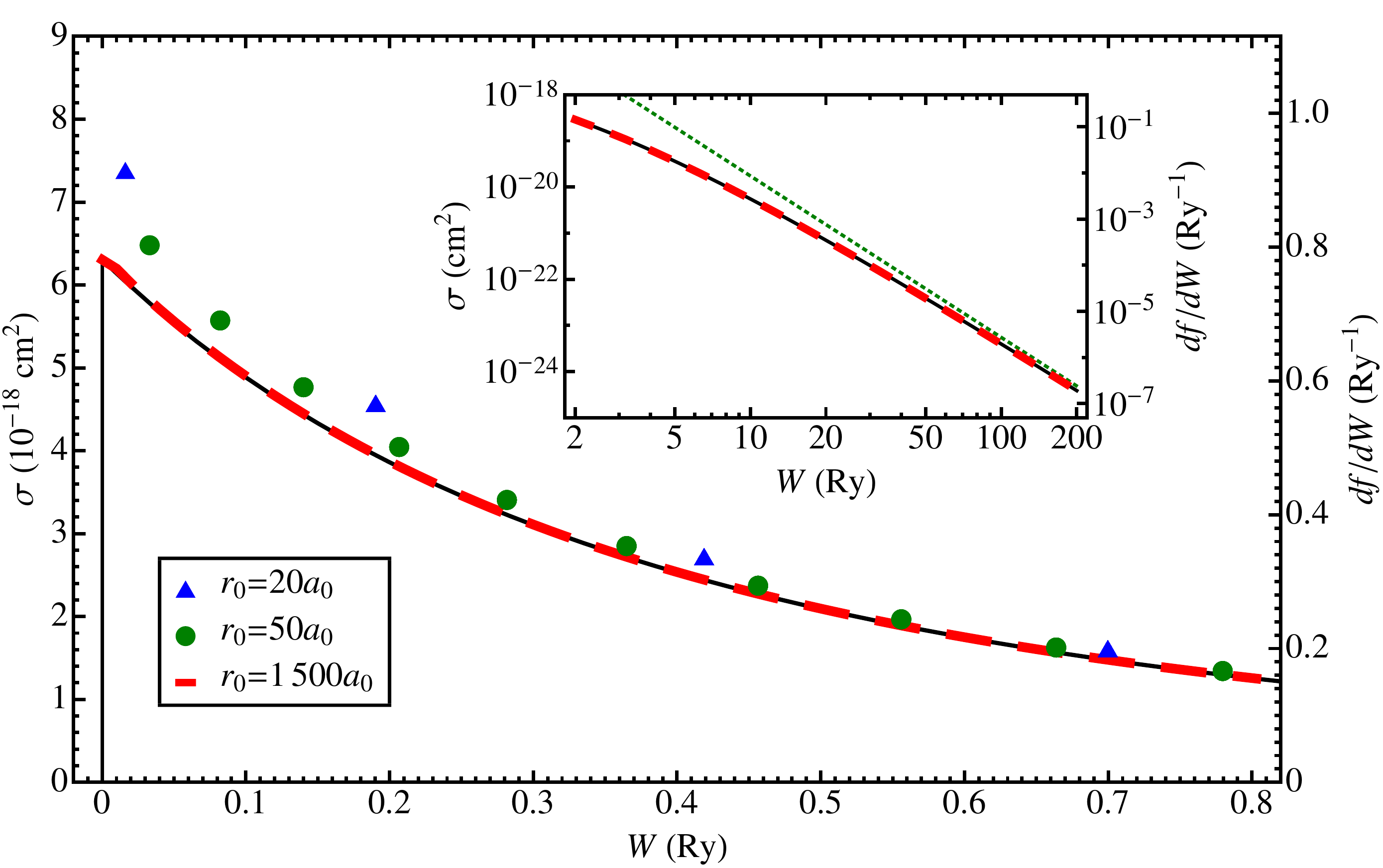}
 \caption[Photoionization cross-section for \mbox{$1\mathrm{s}\rightarrow W\!\mathrm{p}$}]{The photoionization cross-section for \mbox{$1\mathrm{s}\rightarrow W\!\mathrm{p}$.}  The results from the infinite well approach are shown with triangles for \mbox{$r_0=20\unit{a_0}$,} circles for \mbox{$r_0=50\unit{a_0}$,} and a dashed line for \mbox{$r_0=1500\unit{a_0}$,} and the solid line is the exact calculation.  The last two cases are shown in the inset for large energies, along with the dotted line showing the Born approximation.}
 \label{fig:oscstrengthplot1sWp}
\end{figure}

We have calculated the photoionization cross-sections for the hydrogen atom using the infinite well approach for \mbox{$1\mathrm{s}\rightarrow W\!\mathrm{p}$} for a variety of values of $r_0$ and show in \autoref{fig:oscstrengthplot1sWp} that it converges towards those found using exact calculations from \autoref{sec:btc}, which agree with those from Ref.~\citenum{fano}.
The data at $r_0=20\unit{a_0}$ and $r_0=50\unit{a_0}$ is sparse as the densities of pseudocontinuum states at these radii of confinement are quite low; for theses cases all allowed final energies in the range shown are plotted.  With $r_0=1500\unit{a_0}$, there are sufficiently many states in the energy range shown to produce a smooth curve on the scale of the figure.  The consistency with Fermi's Golden Rule is now apparent, in which the density of states evolves from an inverse energy difference: multiplication of the square of the matrix element by the density of states is mathematically equivalent to instead using an en\-\mbox{ergy- ra}\-ther than \mbox{space-nor}malized continuum wavefunction.  It can be seen that the cross-section converges to the true continuum case from above as $r_0$ increases.  This can be attributed to the fact that while $r\rconf_{W\ell}$ tends to a constant amplitude as $r\rightarrow\infty$ (as shown in \autoref{fig:rplot}), it does so from below; thus at low $r$ the amplitude is slightly reduced from this value.  Therefore when the wavefunction is truncated at finite $r_0$ before reaching its final amplitude, $\rconf_{W\ell}^2$ decreases disproportionately to increases in $r_0$.  This results in the $\mathfrak{D}_{W_{n\ell}\rightarrow W\ell'}$ of \autoref{eq:curlyD} taking slightly too large a value for finite $r_0$.  The inset of \autoref{fig:oscstrengthplot1sWp} shows that this calculation works for a large range of final-state energies, over the entire non-relativistic regime.  At high energies, the cross section can be seen to approach that predicted by the Born approximation \cite{fano}, which for a transition $n\mathrm{s}\rightarrow W\mathrm{p}$ takes the form:
\begin{equation}
 \sigma\xrightarrow{W\rightarrow\infty}\frac{2^9\pi \mathrm{a}_0^3 \mathrm{Ry}^{9/2}}{3c\hbar n^3}W^{-\sfrac{7}{2}}.
\end{equation}
This can be arrived at by treating the continuum state as initially being that of a free particle ($e^{ikx}$), and then taking the Coulombic potential into account through first order perturbation theory.

\begin{figure}[tb]
 \includegraphics[width=\columnwidth]{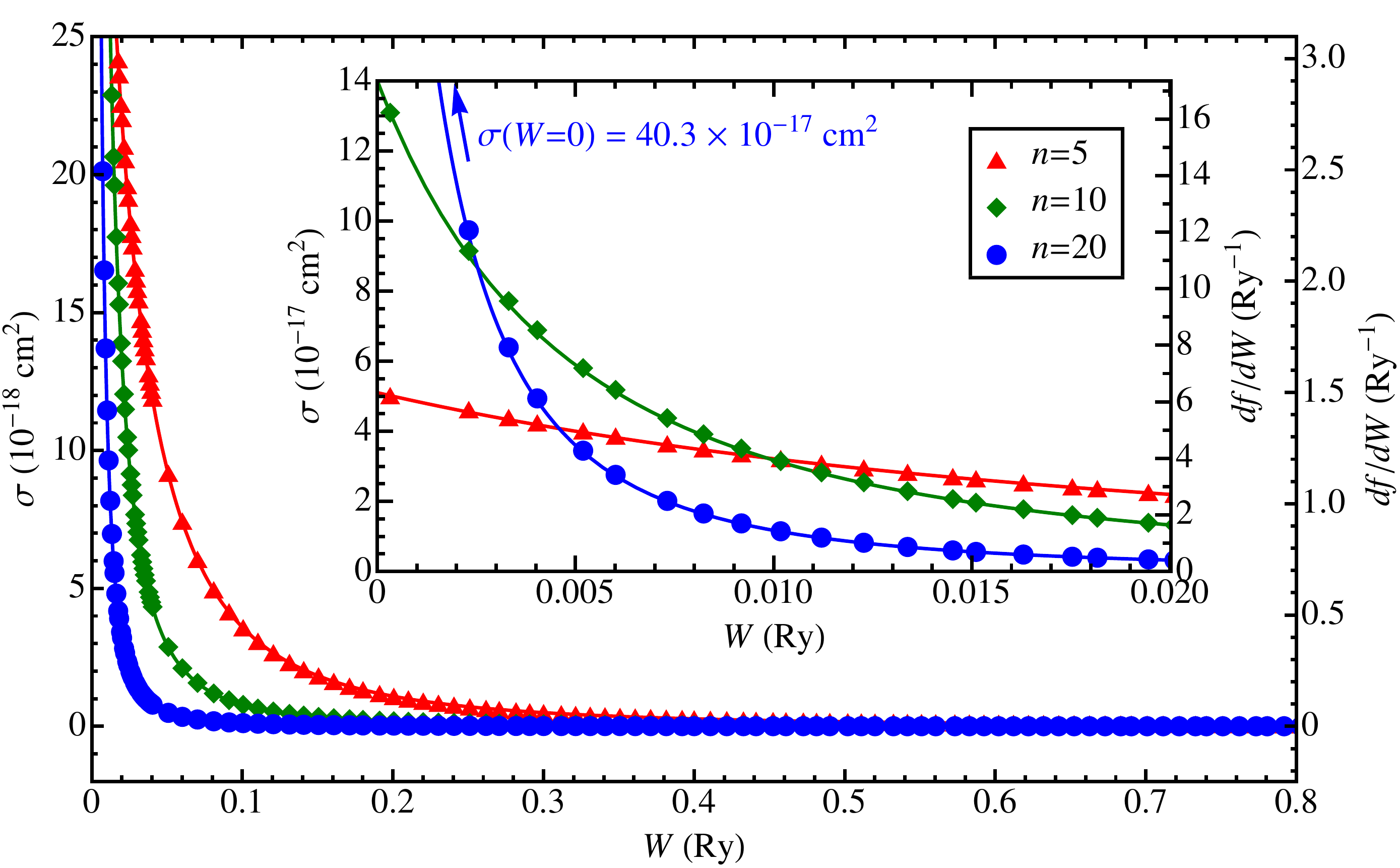}
 \caption[Photoionization cross-sections for \mbox{$\left\lbrace 5,\,10,\,20\right\rbrace\mathrm{s}\rightarrow W\!\mathrm{p}$}]{The photoionization cross-sections for \mbox{$\left\lbrace 5,\,10,\,20\right\rbrace\mathrm{s}\rightarrow W\!\mathrm{p}$.}  The exact calculations are shown as lines, with symbols showing results using the infinite well approach with \mbox{$r_0=1500\unit{a_0}$}. $\mathrm{Ry}$ is the Rydberg unit of energy.}
 \label{fig:crossSection-nsWp}
\end{figure}

An alternative measure of the likelihood of a bound-to-continuum transition is the oscillator strength distribution \cite{gallagher}, $\frac{df}{dW}$, which is related to the cross-section, by
\begin{equation}
 \frac{df}{dW}=\frac{c\hbar}{8\mathrm{a}_0^3\mathrm{Ry}^2\pi^2}\sigma ,
\end{equation}
and for completeness these values are given in Figures \ref{fig:oscstrengthplot1sWp} and \ref{fig:crossSection-nsWp} as well.

We also show in \autoref{fig:crossSection-nsWp} a plot to illustrate that this approach works for states with a range of initial $n$.  Higher $n$ results in a greater cross-section for low ejection energies, and a lesser cross-section for high ejection energies.  This relates to the general observation \cite{merkt} that high $n$ (Rydberg) states are strong absorbers of long-wavelength radiation.  The peak cross section occurs at ionization threshold; for $n\mathrm{s}\rightarrow\br{W{=}0}\!\mathrm{p}$, this simplifies to
\begin{equation}
 \sigma\fbr{W=0}=\frac{2^7n^5\pi^2\mathrm{a}_0^2}{411}\!\br{\sum\limits_{m=0}^{\infty}\!\frac{2^m\br{1{-}n}^{\br{m}} L_{-5-m}^{\br{3}}\fbr{-2n}}{m!}}^2,
\end{equation}
where $L_n^{\br{\alpha}}\br{x}$ is a generalized Laguerre polynomial, and
\begin{equation}
 \br{1-n}^{\br{m}}=\br{1-n}\br{2-n}\cdots\br{m-n}
\end{equation}
is the Pochhammer function, which becomes zero for any $m\ge n$, forcing the sum to terminate.

\section{Conclusions}
We have presented two methods for calculating bound-to-continuum cross-sections: the first employing energy-normalization and the second, box normalization, and demonstrated convergence between them, using photoionization of the hydrogen atom as a concrete example.  We thereby elucidate some quantitative aspects of Fermi's Golden Rule and show the product of the density of states and a space-normalized pseudo-continuum wavefunction to be mathematically equivalent to using only an energy-normalized continuum wavefunction.

\chapter{Cross sections for photo-dissociation of \texorpdfstring{\ce{Li+...I-}}{Li+...I-}}\label{ch:ionpair}
\section{Introduction}
We wish to examine photo-dissociation of long-range, weakly bound heavy Rydberg systems.  We therefore apply our box normalization method to the case of photo-dissociation of heavy Rydberg \ce{Li+...I- + $\hbar\omega$ -> Li+ + I-} where unnormalized wavefunctions are not analytically known, but for which a potential energy curve \emph{is} known from Pan and Mies \cite{pan}.  We first discover numerical wavefunctions for the $J=0$, $0\le v\le 20$ states for which Pan and Mies have reported energies, and find agreement in allowed energies with them in all but their last (\nth{6}) significant figure.  We then use the same method to discover wavefunctions for a Rydberg state and a range of pseudocontinuum states, to which we can then apply the box normalization method outlined in \autoref{sec:wellH} to obtain a photo-dissociation cross-section, $\sigma\fbr{\omega}$.

\section{Methods}
We use the shooting method \cite{press-shooting} with the fourth-order Runge-Kutta method \cite{press-runge} to calculate numerical bound and pseudo-continuum wavefunctions for the system in a box.  Normalization to unity of the bound states gives the in-box bound radial wavefunction $\mathfrak{R}_{vJ}$ and box normalization of the pseudo continuum states as described in section \autoref{sec:wellH} gives box normalized pseudocontinuum radial wavefunctions $\sqrt{\rho\fbr{W}}\mathfrak{R}_{WJ}$.

\begin{figure}[tb]
 \includegraphics[width=\columnwidth]{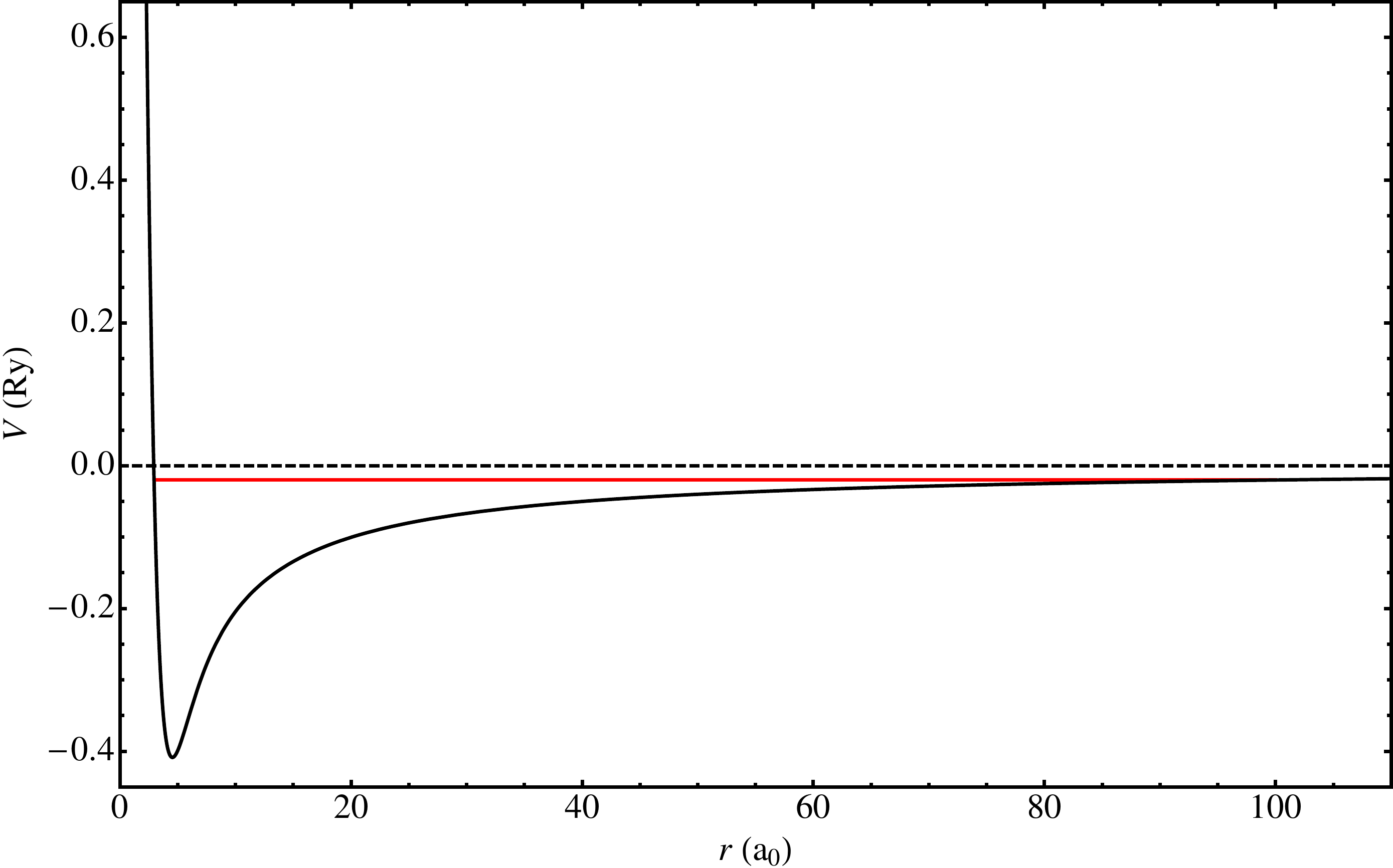}
 \caption[Potential energy for \ce{Li+I-}]{The potential energy, $V$, of \ce{Li+I-} with $J=0$ as given by Pan and Mies \cite{pan} is shown as a function of internuclear separation, $r$.  The red line denotes the energy of the chosen heavy Rydberg \ce{Li+...I-} state with $v=597$.}
 \label{fig:LiIPotential}
\end{figure}

The Rittner potential energy function given by Pan and Mies \cite{pan}\footnote{It should be noted that there is a typographical error in Pan and Mies' Equation (1b), where they have dropped the $-2 \alpha_M \alpha_X r^{-7}$ term found in Ref.~\citenum{brumer}, though appears that they used the correct equation in their calculations.},
\begin{align}
\begin{split}\label{eq:panmies}
 V\fbr{r}&=-\frac{Z}{r}+\frac{J\br{J+1}}{2mr^2}+U_0\fbr{r}\\
 U_0\fbr{r}&\approx\left[A+\br{\frac{B}{r}}^8\right]e^{-\sfrac{r}{a}}-\frac{C}{r^6}-\frac{2 \alpha_M \alpha_X}{r^7},
 \end{split}
\end{align} 
is shown in \autoref{fig:LiIPotential}
for $J=0$.  The values of the constants used in \autoref{eq:panmies} can be found in \autoref{tab:potconstants}.
\begin{table}
\centering
\begin{tabular}{ l l l l }
\toprule
  $A=77.32\unit{Ry}$ & $B=2.508\unit{Ry^{\sfrac{1}{8}}}\unit{a_0}$ & $C=2.75\unit{Ry }\unit{a_0^6}$ & $m=11997m_e$\\
  $a=0.7155\unit{a_0}$ & $\alpha_M=0.20\unit{a_0^3}$ & $\alpha_X=43.40{a_0^3}$ & $Z=1$\\
  \bottomrule
\end{tabular}
\caption{Rittner potential parameters for LiI}
\label{tab:potconstants}
\end{table}
The terms of the Rittner potential may be understood as follows: the $r^{-1}$ term is the Coulombic attraction; the $r^{-2}$ term is a centrifugal term associated with the orbital angular momentum of the two nuclei; the $e^{-r/a}$ term is due to repulsion from electron cloud overlap; the $r^{-8}e^{-r/a}$ phenomological term is added to remove the effects of the longer-range terms at short range; the $r^{-6}$ term is the van der Waals attraction; and the $r^{-7}$ term was suggested by Rittner \cite{rittner}, prompted by disagreement between experimentally measured dipole moments and those calculated using a point charge model.

To mirror calculations performed in previous sections for the hydrogen atom: $ns\rightarrow Wp$, we initially choose a heavy Rydberg state with angular momentum, $J$, equal to zero and therefore pseudocontinuum states with $J=1$. We arbitrarily choose the energy of the Rydberg state to be approximately equal to the potential energy at $r=100\unit{a_o}$, resulting in a vibrational quantum number of $v=597$ (or principal quantum number, $n=v+J+1=598$), and binding energy of $-0.01997\unit{Ry}$. This state is sufficiently long range to be a typical Rydberg state with exaggerated properties, but which begins to decay in $r$ sufficiently soon that calculations are not overly lengthy. The accuracy of the results depends on the size of the box in which the system is placed and over which integrations take place. The box must therefore extend sufficiently beyond the turning point of the Rydberg state chosen.  This will be seen in \autoref{sec:liiresults} below.

We solve for the wavefunctions of the Rydberg state and of the pseudo-continuum states over a range $1.5\unit{a_0}<r<r_0$, with the lower bound chosen to avoid the infinite potential energy at $r=0$, but lying sufficiently far into the region of exponential decay for all energies studied that the wavefunction is close enough to zero that all values $\Psi\fbr{r<1.5\unit{a_0}}$ can reasonably be considered to be zero.

The shooting method is applied: starting at $r=1.5\unit{a_0}$ for a guessed energy and arbitrary initial value and slope, and then using the Runge-Kutta algorithm to iteratively solve the Schr\"odinger equation.  The initial value is arbitrary since the wavefunction will later be normalized, and the initial slope is arbitrary since the method is started in a region of exponential decay, and the result relies only very weakly on the initial slope provided. To test this last point, initial slopes varying by 2 orders of magnitude were chosen, and the normalized results varied by no more than 0.4\% in wavefunction magnitude at any point. This level of deviation applied only near the chosen starting point, and dropped of quickly away from that point.  As we always choose starting points in regions of exponential decay where the wavefunction is approximately zero, this deviation has very little effect.

The heavy Rydberg state calculated begins and ends in regions of exponential decay.  Since the guessed energy is never exactly correct, in the region of decay opposite that in which the shooting method is started, the function shoots off to $\pm\infty$, with the sign dependent on whether the energy is too high or low (and the number of nodes in the wavefunction).  By choosing energies near which we'd like to find wavefunctions and counting nodes in the resultant (incorrect) wavefunctions to ensure that two chosen energies surround only a single eigenenergy, a window $W_\mathrm{min}\to W_\mathrm{max}$ is created within which the correct energy must lie.  A binary search can be performed by applying the shooting method at $\sfrac{\br{W_\mathrm{min}+W_\mathrm{max}}}{2}$ and setting either $W_\mathrm{min}$ or $W_\mathrm{max}$ to this new value, depending on the sign of the calculated wave function at large $r$.  This is iterated, reducing the window size to within an acceptable error.  This wavefunction will still shoot off to $\pm\infty$ at large $r$, but a method for producing a clean wavefunction is described below.

After calculating the wavefunction at the discovered energy starting at low $r$, a second calculation is performed, starting at high $r$ and shooting to the left.  For sufficiently precise chosen energies these functions will be phase locked, having the same functional form except where they begin to shoot off to $\pm\infty$ in the region of exponential decay opposite their starting points.  The final clean wavefunction is produced by choosing an $r_\mathrm{cut}$ somewhere in the middle, and taking for $r<r_\mathrm{cut}$ the values calculated starting at low $r$, and for $r>r_\mathrm{cut}$, the values calculated starting at high $r$.  Though the two calculations will be phase locked, they will not necessarily have the same amplitude, so one side is scaled to achieve continuous values in the final wavefunction.  Which side is scaled is unimportant, as the wavefunction will be normalized before use in other calculations.  In order to minimize the error in scaling, $r_\mathrm{cut}$ is chosen to be at a local maximum (or minimum) of the wavefunction.  The first derivative will also be continous since the two calculations are phase locked. The final result is a clean wavefunction for the heavy Rydberg state as it avoids using any values in a region of decay calculated at the end of a shooting method which erroneously shoot off to $\pm\infty$.

To determine the wavefunctions of the pseudocontinuum states of \ce{Li+ + I-}, an energy window surrounding only one state is determined as above for the Rydberg state. In this case, a full binary search is not required, since near the hard wall at $r_0$ these states are quite sinusoidal and therefore more easily predictable.  Knowledge of the end points and slopes of the (incorrect) wavefunctions associated with the maximum and minimum energies is used to algorithmically estimate the correct energy.  The algorithm, which may be found in the program in \autoref{app:code}, is often sufficiently accurate to determine the correct energy within the required precision in one step.

Since the angular portion of the wavefunction of a diatomic molecule is the same spherical harmonic \cite{bernath169}, $Y_J^M$, as for hydrogen, \autoref{eq:sigmacont} need only change in its labelling to find the cross-sections for dissociation of ion pairs:
\begin{equation}
 \sigma\fbr{\omega}=\lim_{r_0\to\infty}\frac{4\pi^2\omega \mathrm{a}_0 2\mathrm{Ry}}{3c}\frac{J_\mathrm{max}}{2J+1}\left|\mathfrak{D}_{vJ\rightarrow WJ'}\right|^2,
\end{equation}
where $J_\mathrm{max}$ is the greater of $J$ and $J'$, $\hbar\omega$ is the difference in energy between the upper and lower states, and \autoref{eq:curlyD} is relabelled to produce
\begin{equation}
 \mathfrak{D}_{vJ\rightarrow WJ'}=\int\limits_0^{\infty}{\sqrt{\rho\fbr{W}}\mathfrak{R}_{WJ'}^\mathrm{\,en}}^{\hspace{-0.5em}*}\,r\,\mathfrak{R}_{vJ}\,r^2\,dr.
\end{equation} 

We also calculate cross sections for an initial Rydberg state with an energy close to that of the $\ket{v=597,\,J=0}$ state used above, but with the highest possible angular momentum at that energy: $\ket{v=0,\,J=774}$ ($n=775$).  We employ the same process as used above to calculate this cross section for comparison of high angular momentum states.

\section{Results and Discussion}\label{sec:liiresults}
\begin{figure}[tb]
 \includegraphics[width=\columnwidth]{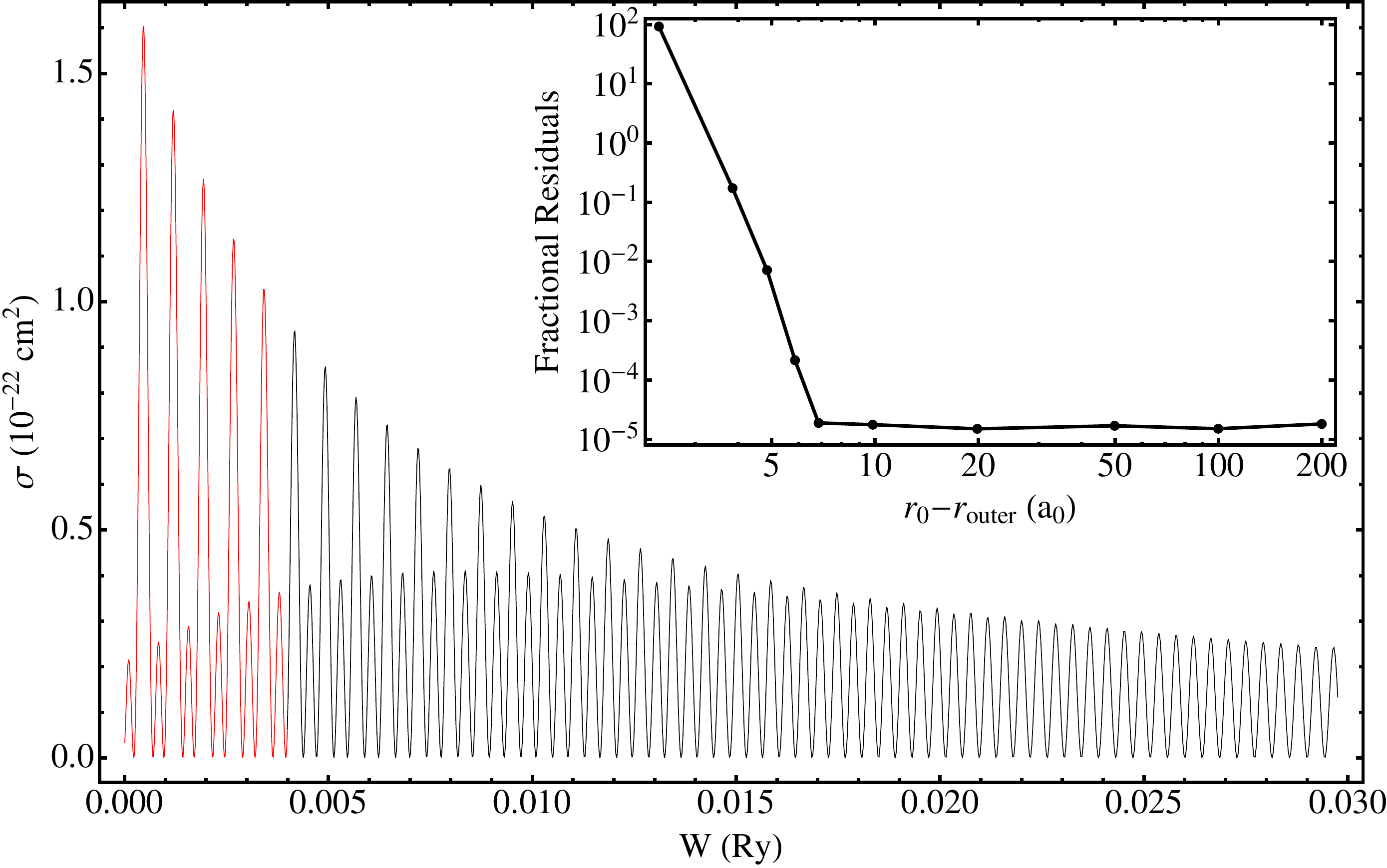}
 \caption[Cross-section for \ce{Li+...I-} $\ket{v=597,J=0}\rightarrow\ket{W,J=1}$]{Calculated photo-dissociation cross-section of \ce{Li+...I-} $\ket{v=597,J=0}$ $\rightarrow$ $\ket{W,J=1}$ using $r_0=400\unit{a_0}$.  The inset shows fractional residuals: the scaled RMS of the difference between the cross-section calculated using $r_0$ as shown and that using $r_0=400\unit{a_0}$ over the highlighted range $0<W<0.004\unit{Ry}$, with the scaling at all $W$ values defined as a division by the maximum value of $\sigma$ in that range. The outer turning point, $r_\mathrm{outer}$, is approximately $100\unit{a_0}$.}
 \label{fig:LiI-crossSection}
\end{figure}

The cross-section for photo-disociation of \ce{Li+...I-} $\ket{v=597,J=0}\rightarrow\ket{W,J=1}$ using a spherical well with $r_0=400\unit{a_0}$ is shown in \autoref{fig:LiI-crossSection}.
The region $0<W<0.004\unit{Ry}$ is highlighted since the fractional residuals shown in the inset were calculated over this range.  They were generated by first creating a third-order interpolation of $\sigma_{r_0=400\unit{a_0}}\fbr{W}$ in Mathematica{\textregistered} since changing $r_0$ changes the discrete energies at which $\sigma$ is calculable.  Then for the various $r_0$ plotted in the inset, states with energies in the range $0<W<0.004\unit{Ry}$ were found and the associated cross-sections calculated.  The fractional residual plotted is the scaled root-mean-square value
\begin{equation}
 \frac{1}{\mathrm{max}\fbr{\sigma_{400\unit{a_0}}\fbr{W_i}}}\sqrt{\frac{1}{n}\sum\limits_{i=1}^n \br{\sigma_{r_0}\fbr{W_i}-\sigma_{400\unit{a_0}}^\mathrm{interp.}\fbr{W_i}}^2},
\end{equation}
where $i$ indexes the energies of the $n$ states in the chosen range.  This is plotted in the inset on a log-log plot with $r_\mathrm{outer}\approx 100\unit{a_0}$, showing that $r_0=400\unit{a_0}$ is far larger than it need be to get an accurate cross-section as the calculated cross-sections asymptote in $r_0$ to their final values very quickly for $r_0>r_\mathrm{outer}$.

The oscillatory nature of the cross-section is surprising as it is not something we observed for hydrogen. The oscillations have, however, remained consistent through the modification of numerical parameters for the radius of confinement and the precision to which energy levels were calculated; as such we believe the oscillations to be physical.  It is also worth noting that though the cross section oscillates, it does still lie within a decaying envelope as indeed it must due to increasing cancellation due to increasing oscillation about zero of the product of states being integrated.

\begin{figure}[tb]
\includegraphics[width=\columnwidth]{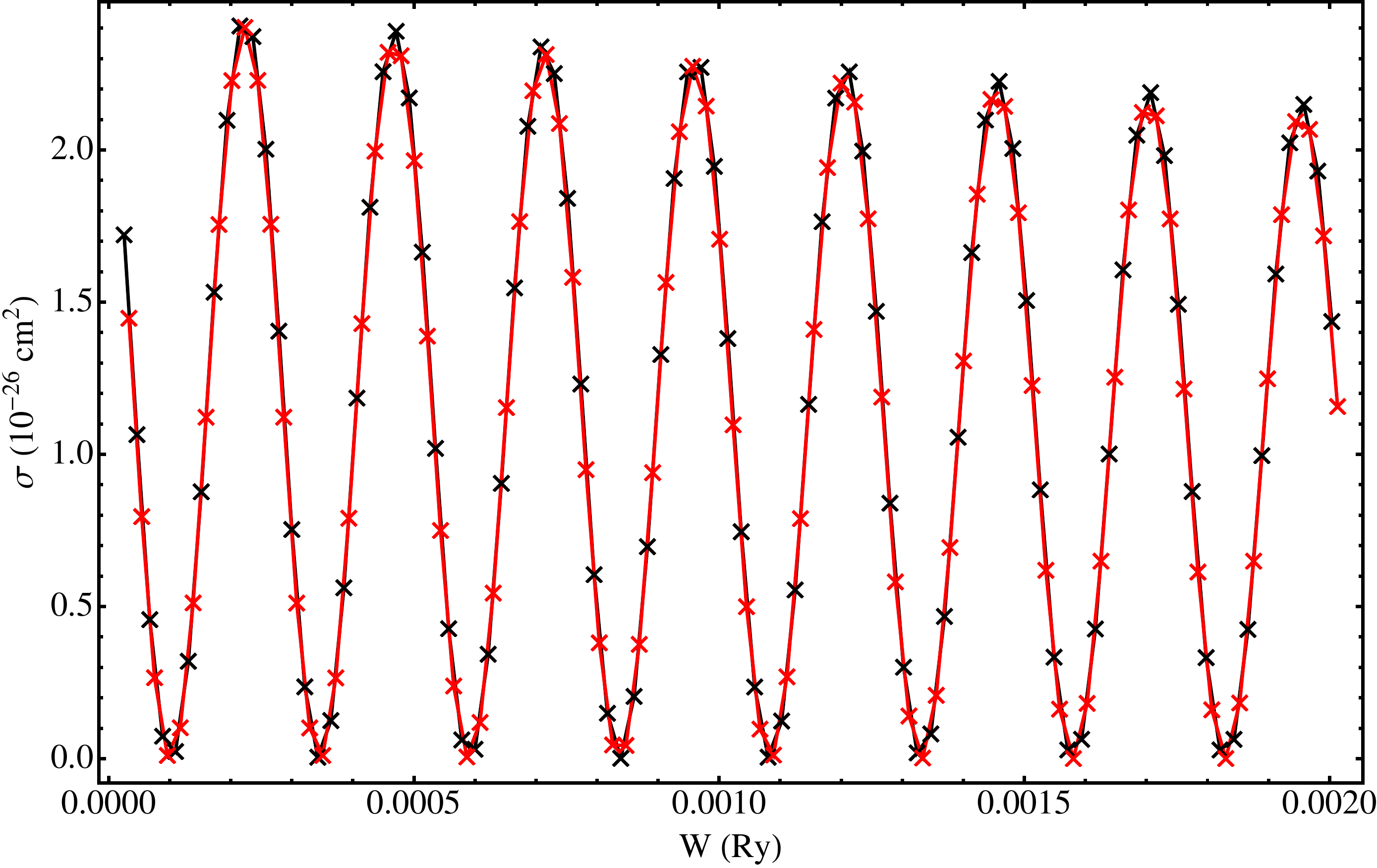}
 \caption[Cross-section for \ce{Li+...I-} $\ket{v=0,J=774}\rightarrow\ket{W,J\in\{773,775\}}$]{Calculated photo-dissociation cross-section of \ce{Li+...I-} $\ket{v=0,J=774}$ $\rightarrow$ $\ket{W,J=775}$ in black and $\ket{v=0,J=774}\rightarrow\ket{W,J=773}$ in red using $r_0=300\unit{a_0}$ in both cases.}
 \label{fig:LiI-highJcrossSection}
\end{figure}

For comparison, in \autoref{fig:LiI-highJcrossSection}
cross sections are shown for $\ket{v=0,J=774}\rightarrow\ket{W,J=775}$ and $\ket{v=0,J=774}\rightarrow\ket{W,J=773}$ using $r_0=300\unit{a_0}$.  RMS fractional residuals (calculated as above over the plotted range) between the cross-sections shown and those calculated with $r_0=200\unit{a_0}$ are $\approx 5\times 10^{-4}$ in both cases.  That the two cross-sections plotted in \autoref{fig:LiI-highJcrossSection} are so similar is not surprising as the effective potential from which the upper states are calculated differs only in the $r^{-2}$ term for which the coefficients differ by a mere 0.5\%. The only other difference in the calculation of the cross section is the pre-factor $J_\mathrm{max}$ which differs by 0.1\% between the two.  We do resolve a difference between the two cross-sections, however, as the cross-calculated RMS fractional residuals (calculated between the two cross sections) is $8\times 10^{-3}$, an order of magnitude greater than that of each cross section using $r_0=300\unit{a_0}$ with itself calculated at $r_0=200\unit{a_0}$.

\section{Conclusions}
We have calculated cross-sections for photo-dissociation of various states of \ce{Li+...I-} which are, to the best of our knowledge, the first heavy Rydberg photo-dissociation cross-sections presented. In the case of highest possible angular momentum for the chosen energy, we find that the oscillations in $\sigma\fbr{W}$ have a slightly higher frequency than those for the corresponding zero angular momentum case, and that the amplitude of the cross-section is 4 orders of magnitude less.

The oscillatory nature of the cross-sections for each initial state considered has remained consistent through modifying several numerical parameters and as such we believe the oscillations to be physical.  This suggests the possibility of selective control using light which cannot dissociate the ion pair but only force a particular desired downward transition.

This work lays excellent groundwork for minor adjustments to be made to the parameters of the program found in \autoref{app:code} in order to calculate cross-sections for \ce{Li+...Li-}, or indeed any other ion pair.

\part{Wavemeter redesign and testing}\label{part:wavemeter}

\chapter{Literature review: Wavelength measurement}
Measuring the wavelength (or, equivalently, frequency) of electromagnetic radiation can be performed in a variety of ways.  With very coarse accuracy within the visible range, our eyes are capable of determining wavelengths in the form of colours.  In order to make more precise measures of wavelength we must turn to scientific instruments.  One such option is a diffraction grating spectrometer \cite{pedrotti} in which the frequency of light is angularly dispersed due to being diffracted.  A scanning Fabry-P\'erot optical spectrum analyser (OSA) \cite{hollas} can get considerably better resolution, but cannot produce absolute frequency measurements as it provides a response modulo the free spectral range (FSR), determined by the length of the cavity and the wavelength being examined.  Multiple scanning Fabry-P\'erot OSAs each with a different FSR may be used to recover the absolute frequency \cite{fischer}, but this can quickly become expensive, and is specific to a small wavelength range due to the high reflectivity coatings necessary for high finesse etalons.  A Fizeau wavemeter built by Morris \emph{et al.} \cite{morris} employs a Fizeau wedge interferometer which they built and tested, and a photo-diode array.  

Another option, and the one we choose, is the Michelson interferometer wavemeter which uses a reference laser with known wavelength to determine the change in optical path length also experienced by the laser beam under test. The original such design was published in 1976 by Kowalski \emph{et al.} \cite{kowalski}, and proved accurate to 6 parts in $10^8$.  Two years later Kowalski published another paper with other collaborators \cite{kowalski1978} in which they modified their original design to improve the accuracy by a factor of 6.  More recently, Fox \emph{et al.} \cite{fox} provide a slightly less precise but also less expensive wavemeter. They also introduce a signal processing increment/decrement scheme which we also employ for digitally determining the wavelength that avoids performing a division by instead choosing to cease counting fringes at the appropriate moment, as will be described in \autoref{sec:sigproc}.  This is useful as division is a lengthy operation for the inexpensive microcontrollers we use; on our particular device division is not implemented as a built in instruction which means a subroutine would have been required which would have performed division with multiple instructions involving many subtractions.  On those devices where division is implemented, a division between two bytes takes many instruction cycles ($\sim 20$), while most operations take only one.  The ultimate measurement of optical wavelength or frequency, naturally, is not through physical scientific instruments in isolation, but by calibrating against primary elemental standards through such methods as spectroscopy of atoms or molecules \cite{libbrecht}.

\longchapter{Redesigning a wavemeter for increased reliability and ease of use}{Wavemeter development}\label{ch:wavemeter}

\section{Introduction}
One of the steps in setting up the magneto optical trap to study cold collisions in lithium will be to tune three external cavity diode lasers close to the D2 line of Lithium ($446\,810\unit{GHz}$).  One laser will be locked to a saturated absorption spectrum \cite{demtroder} of the D2 line, and the other two lasers will be offset locked \cite{ritt} to the first, tuned to cooling transitions in lithium.  Since these diode lasers can be tuned over hundreds of gigahertz in frequency by changing the angle of the grating which creates the external cavity, searching blindly for a transition with a width on the order of a few megahertz is entirely impractical.  To that end, a wavemeter was developed by me (M. Ugray), and three other co-authors \cite{ugray2006} in the lab several years ago to allow for coarse tuning of the lasers to within no more than a few gigahertz of laser frequency.  While that wavemeter has provided accurate results and been used to tune to the D2 line \cite{philippson}, it has always been time consuming and finicky to align properly, and its continued use was taking time away from the experiment.

\begin{figure}
\includegraphics[width=\columnwidth]{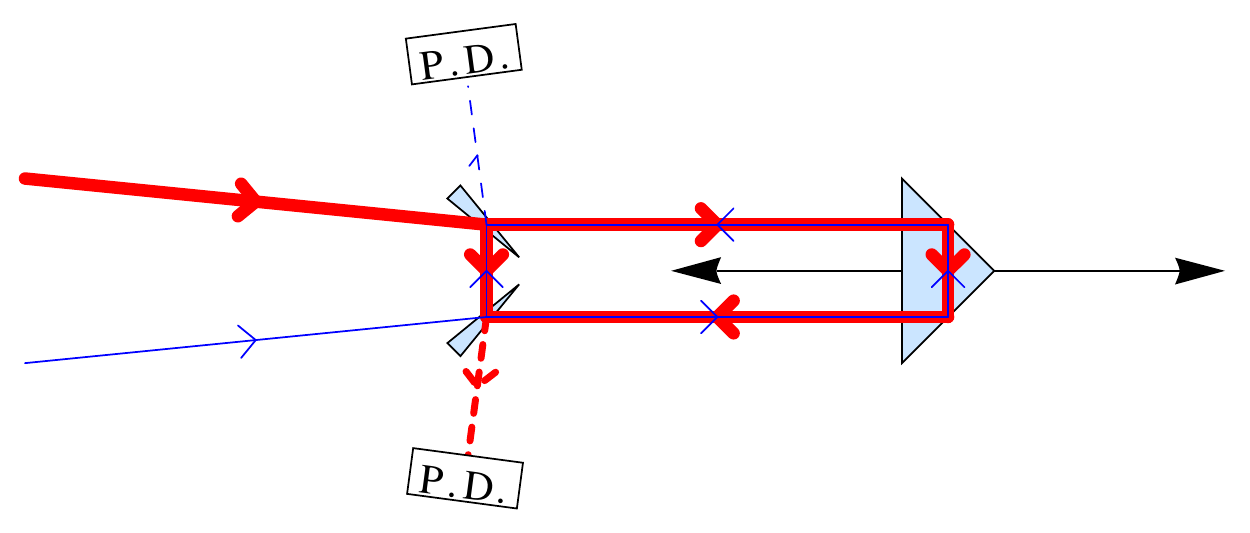}
 \caption[Previous wavemeter optical path layout]{The optical path layout of the previous wavemeter.  The red path indicates the reference HeNe beam of known wavelength while the blue path indicates the beam of the laser under test, of unknown wavelength.  Dashed lines indicate the presence of interference fringes which are observed on the photodiodes marked P.D.}
 \label{fig:old-wavemeter-diagram}
\end{figure}

The initial design of the wavemeter, described in Ref.~\citenum{ugray2006}, can be divided into three sub-parts: the optical train, signal digitization, and digital signal processing.  The optical train, shown in \autoref{fig:old-wavemeter-diagram},
employed a linearly travelling Michelson interferometer design.  It incorporated a triangular prism retroreflector attached to a cart moving back and forth along a linear track, and a stabilized HeNe reference laser of known wavelength.  The moving retroreflector caused a changing \emph{optical path length difference} (OPLD) between the two paths, resulting in time resolved interference fringes of each laser with itself.  The sinusoidal interference signal at each photodiode goes through a full period for each change in the OPLD equal to the vacuum wavelenth, $\lambda$.  As the retroreflector moved through a distance $\Delta L$, the change in OPLD for each beam was $2n\fbr{\lambda}\Delta L$, where $n\fbr{\lambda}$ is the index of refraction at the wavelength of the beam in question.  If the bright fringes at the photodiode are counted as the retroreflector travels, then
\begin{equation}\label{eq:michelson}
N\lambda=2n\fbr{\lambda}\Delta L
\end{equation}
with $N$ being the number of bright fringes counted as the retroreflector moves.  Applying this equation for both beams and adding subscripts to differentiate, and with the understanding that for visible wavelengths in air $n\fbr{\lambda_\mathrm{HeNe}}\approx n\fbr{\lambda_\mathrm{unknown}}$, we solve each equation for the common $\Delta L$, set them equal to each other and arrive at
\begin{equation}
\lambda_\mathrm{unknown}=\frac{N_\mathrm{HeNe}}{N_\mathrm{unknown}}\lambda_\mathrm{HeNe},
\end{equation}
or, equivalently for frequency, $\nu$,
\begin{equation}\label{eq:funknown}
\nu_\mathrm{unknown}=\frac{N_\mathrm{unknown}}{N_\mathrm{HeNe}}\nu_\mathrm{HeNe},
\end{equation}

\begin{figure}[t]
\includegraphics[width=\columnwidth]{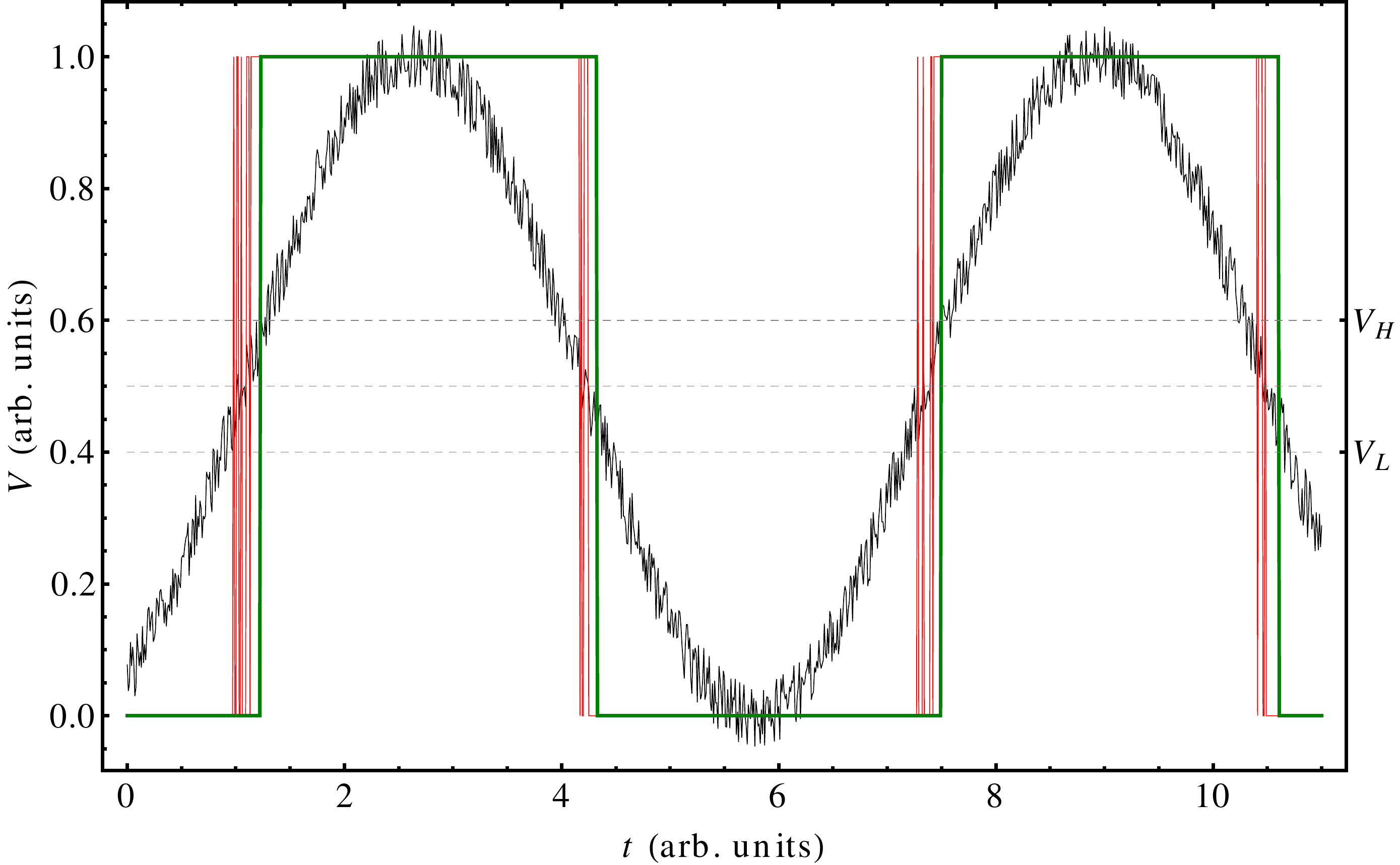}
 \caption[Plot of Schmitt trigger hysteresis]{A plot showing an analogue signal (black), a na\"ive digitization (red) giving high for any value of the analogue signal greater than $0.5$ and low otherwise, and a Schmitt triggered digitization (green).}
 \label{fig:schmitt-hysteresis}
\end{figure}

Since many of these fringes must be calculated, it is necessary to process them using electronics.  The next step, therefore, is to digitize the analogue photodiode signals.  This is done using a Schmitt trigger, a device which digitizes an analogue signal with added hysteresis to avoid counting erroneous peaks due to the noise that will inevitably be a part of any analogue signal.  To illustrate the way it works, a sinusoidal signal with added random noise was generated on a computer, and is shown in \autoref{fig:schmitt-hysteresis}, 
with two digitized outputs, one using the concept of the Schmitt trigger, and the other a na\"ive digitization in which the output generated is high for any input above a threshold value of 0.5 in this example, and low otherwise. The output of a Schmitt trigger, on the other hand, remains high until the analogue input goes below a threshold $V_L$, and then remains low until the analogue signal goes above a second threshold $V_H$.  Having two separate threshold values creates a hysteresis window that eliminates the effect of noise: the Schmitt trigger output shows two maxima, the correct response, while the na\"ive digitization shows 13 maxima due to multiple triggerings on noise around threshold.  The Schmitt triggers we use in our wavemeter digitize the photodiode signals; once digitized, the signals must be processed.

Processing of the digitized signals is performed by a programmable microcontroller.  This choice was made since they are robust, inexpensive, and small devices compared with PCs running an operating system outside of our direct control.

The previous wavemeter design had a few problems which have been corrected through redesigning the wavemeter.  First, aligning the wavemeter was quite difficult as although it could be aligned well with the prism of the travelling Michelson at one end of the track, when the prism was moved to the far end of the track, it would be out of alignment due to a lack of straightness of the track.  A compromise was required in which the wavemeter was intentionally slightly misaligned for all track positions, trying to optimize the minimum alignment, wherever that might be.  Second, the mechanism for pulling the cart was such that there were extended periods of time during which the cart was stationary at the ends of the track, giving no updates to the wavelength measurement.  Third, the two Schmitt triggers used to count fringes from the reference and unknown laser beams were built on a noisy board, for which a change of state of one Schmitt trigger occasionally caused the second Schmitt trigger to erroneously also change state.  In the following sections the modifications which fixed these problems are described.

\section{Straightening the track}
The first method used to attempt to improve the ease of use of the wavemeter was to try to straighten the track on which the corner prism travelled.  In attempting this, a number of problems were encountered which encouraged us to find an alternate method.

The first problem was that bolting the track to posts on the optical bench as under normal operation caused the track to bend, and bolting it loosely or not at all would allow the track itself to move under pressure from the driving mechanism of the cart which carried the corner prism along the track.  Together, these meant that any straightness measurements and adjustments would have to be done \textit{in situ}, after the track had been bolted to the bench.

Measuring the straightness of the track \textit{in situ} was also non-trivial.  We choose axes such that the y-direction is the nominally horizontal direction of travel of the cart on the track, and the z-direction lies in a vertical plane.  If we let $\br{x\fbr{y},\,y,\,z\fbr{y}}$ be points along the path of the track for all $y$ between zero and the length of the track, $L$, then the desired case of a straight track is that where
\begin{equation}
\frac{d^2x}{dy^2}=\frac{d^2z}{dy^2}=0\quad\forall y,\, 0 \le y \le L
\end{equation}
To measure $z\fbr{y}$ we tried to use a standard dial indicator.  However this measures the height of an object relative to a flat supporting structure and requires the optical bench upon which the track was mounted to be more flat than the track.  In a limited test of flatness of the track and, separately, a section of optical bench equal in length to the track, the dial indicator was run along the length of each one without mounting it in the height gauge.  This was a limited test of flatness as it referenced the height of each point under the tip of the dial indicator to the $5\unit{cm}$ long base of the indicator (usually used for mounting) $5\unit{cm}$ away.  This method would not, therefore, detect a curve of constant radius in either the track or bench.  The readings for the track varied non-linearly within a window $50\unit{\mu m}$ wide, and those for the table varied non-linearly within a window $430\unit{\mu m}$ wide.  Since the track was already insufficiently flat, the table, being an order of magnitude less flat, was hardly a good candidate as a reference surface.  Additionally, the dial indicator would have been ineffective for measuring the horizontal straightness without some flat reference plane with a horizontal normal.

An alternate attempt was made to assess the straightness of the track using a camera mounted on the cart to view an incoming laser beam, but the data transfer proved overly slow.  In conjunction were the problems of having only six bolt holes as sites to shim and apply torque to straighten the track, wiggle between the cart and the track, and the fact that the wavemeter once set up would not be portable.  We determined that straightening the track was not practical.

\section{Using a rocker}

\begin{figure}[tb]
 \includegraphics[width=\columnwidth]{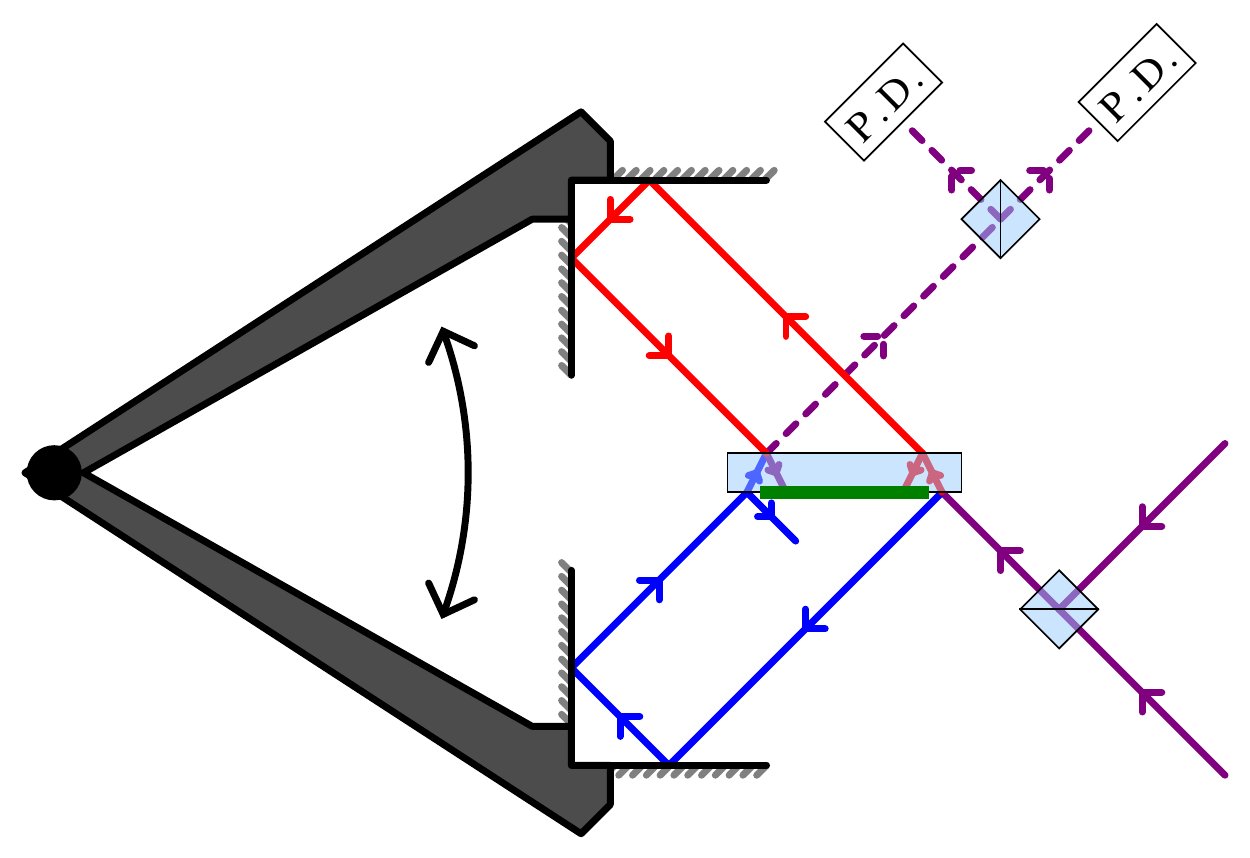}
 \caption[Schematic of the rocker system]{Schematic of the new rocker system, with each 3-orthogonal-mirror retroreflector represented by two mirrors for simplicity.  The two input laser beams are combined by the first polarizing beam splitter and separated towards their respective photo diodes using the second polarizing beam splitter.  The red and blue beams show the two beam paths created by the rocker, with the dashed beam indicating fringes present in the beam due to the changing optical path length difference between the red ad blue paths.  The green line denotes an opaque coating on the beamsplitter, blocking some unwanted beams.}
 \label{fig:wavemeter-diagram}
\end{figure}

It was decided to replace the optical train of the wavemeter with a rocker system from a Bomem FT-IR spectrometer (Michelson MB155) that was disposed of by the Chemistry Department at Trent University.  The rocker uses two corner retroreflectors on armatures that rock together.  The rocking is driven by an electromagnet, which was already integrated into the rocker in the FT-IR spectrometer, attached to a function generator.  Between the two retroreflectors there is a stationary beam splitter.  To achieve the best signal-to-noise at the photodiode, the two paths should have the same power, and as high a power as possible.  The retroreflectors provide close to 100\% reflectivity, so we examine the series of transmissions and reflections at the beamsplitter in each beam path shown shown below in \autoref{fig:wavemeter-diagram}
between the two polarizing beam splitters and find that both have one reflection and 2 transmissions, so maximizing power is the only concern. Assuming no absorption, we set the reflectivity, $R=(1-T)$, with T being the percent transmittance.  The fractional power in each beam is $P=T^2 R=T^2\br{1-T}$, which is maximized for $T=\frac{2}{3}$, $R=\frac{1}{3}$.  This is approximately achieved with a vertically polarized input beam, where we measure $R=0.41$, $T=0.59$, resulting in $P$ at 96\% of the maximum theoretical output power; for measurements for horizontally polarized light, see below.

In an initial configuration, the beam from the laser under test went into the rocker system horizontally at the height of the center of the retroreflectors such that it remained in a constant horizontal plane, while the beam from the reference laser went in above this, and so was translated down by the same amount by each retroreflector, so that the two reference beams were properly recombined, and could be directed to a separate photodiode than that used for the beam under test.  This configuration, however, resulted in the wavemeter reporting frequencies that were about 0.01\% higher than they ought to be, as will be discussed in more detail in \autoref{sec:waveresults} below.  Given that we're looking to use the wavemeter to tune to within a few gigahertz of $446\,810\unit{GHz}$, a 0.01\% error is too high by an order of magnitude.

To correct the error, we tried instead sending both beams along the same path, using a polarizing beamsplitter cube to combine the beams from the reference and laser under test before they entered the rocker system, one vertically polarized and the other horizontally.  The mirrors of the retroreflectors necessarily do not all lie in vertical planes, which means that the polarization of the beams is modified on hitting these mirrors.  However, since the mirrors are all mutually orthogonally oriented, the result of a retroreflection is that a beam regains its initial polarization state, making reseparating the beams after the rocker system possible simply by using a second polarizing beam splitter cube, as shown in \autoref{fig:wavemeter-diagram}.  The fact that one beam is therefore necessarily horizontally polarized going through the (non-polarizing) beam splitter of the rocker system gives that beam $R=0.17 \implies T=0.83$, resulting in 79\% of the theoretical maximum output power.  We chose this beam to come from the reference laser which is entirely devoted to the wavemeter so that the beam under test can have as little power siphoned off to the wavemeter as possible.

\begin{figure}
 \centering
 \includegraphics[width=\columnwidth]{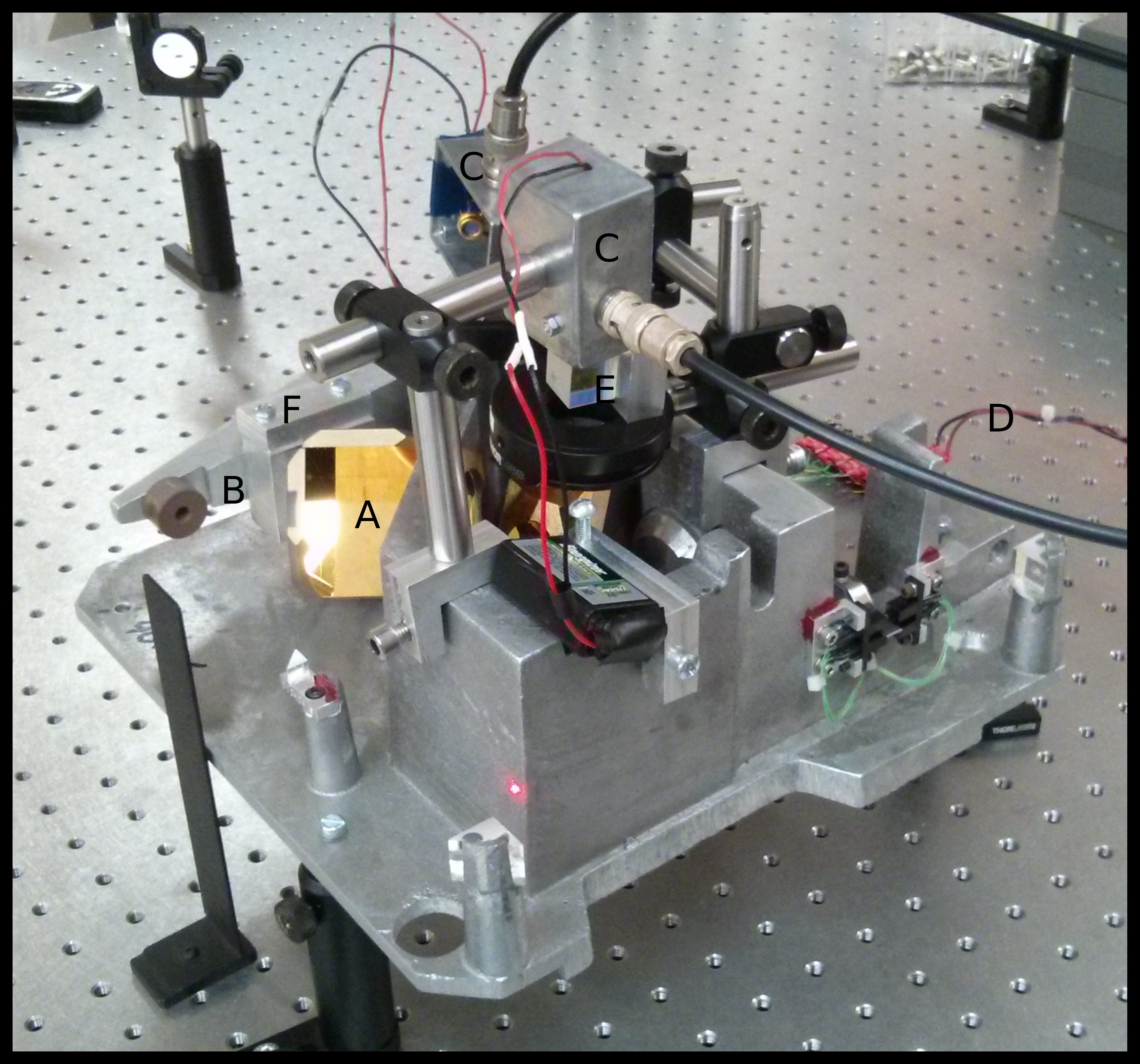}
 \caption[Photograph of the rocker system]{A photograph of the rocker showing the two photodiodes in place above the system. A) retroreflector; B) rocker armature; C) photodiode box; D) electromagnet driving signal; E) polarizing beam splitter; and F) added mass (reduces resonant frequency).}
 \label{fig:rocker}
\end{figure}

The use of a rocker, shown in \autoref{fig:rocker}, has a non-trivial additional benefit: since the nature of the rocker is that it oscillates by a small amount around \emph{zero} path difference, it is not a problem if there are two modes present in the HeNe beam -- the beating of the interference pattern this causes is a long range one which leaves small path differences unaffected.  Since the HeNe laser has a $1.5\unit{GHz}$ wide gain profile, and this is sufficiently stable for the resolution we desire, the only possible reason to have stabilized the laser would be to avoid multi-mode operation.  Since that is not a problem in this case, the wavemeter can be used immediately after it is turned on, without needing to wait for the HeNe laser to warm up and be subsequently stabilized.

\section{Schmitt trigger improvement}

\begin{figure}
 \centering
 \includegraphics[width=0.85\columnwidth]{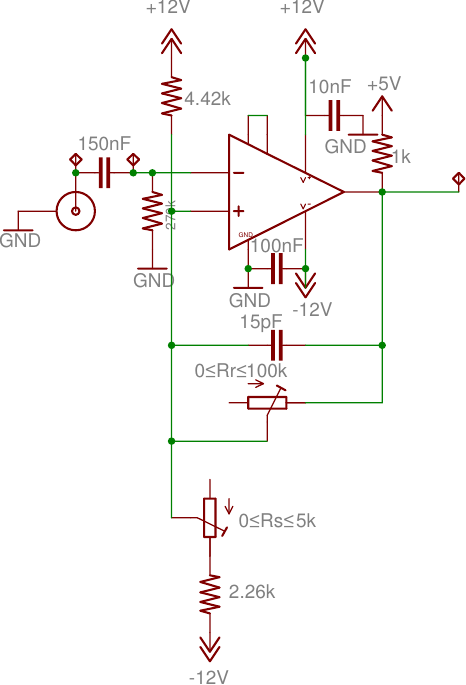}
 \caption[Schmitt trigger circuit diagram]{The circuit diagram for one Schmitt trigger using an LM311 comparator.  The final board had two of these, side by side, sharing power lines.  Trimpots are labelled $R_r$, which mainly controls the range of the trigger window, and $R_s$, which mainly controls the set-point or centre of the trigger window.}
 \label{fig:schmitt}
\end{figure}

The interference signal from each photodiode is an analogue sinusoidal signal in time as the rocker rotates. In order to count the number of maxima with a programmable microcontroller, it is necessary to first reliably digitize the fringe patterns.  This is done with a pair of Schmitt triggers \cite{horowitz}, which avoid triggering on noise by employing hysteresis such that the trigger point where the output switches from high to low depends on the state of the output.

The Schmitt circuits constructed previously \cite{ugray2006} were prone to mis-triggering due to noise, sometimes even cross-triggering due to a trigger from the other Schmitt circuit.  As the Schmitt triggers were built on Veroboard\texttrademark{}, which can be noisy at high frequencies, a printed circuit board with a ground plane was designed (see schematic in \autoref{fig:schmitt}) and constructed to replace the original Schmitt trigger board.

\begin{figure}[tb]
 \includegraphics[width=\columnwidth]{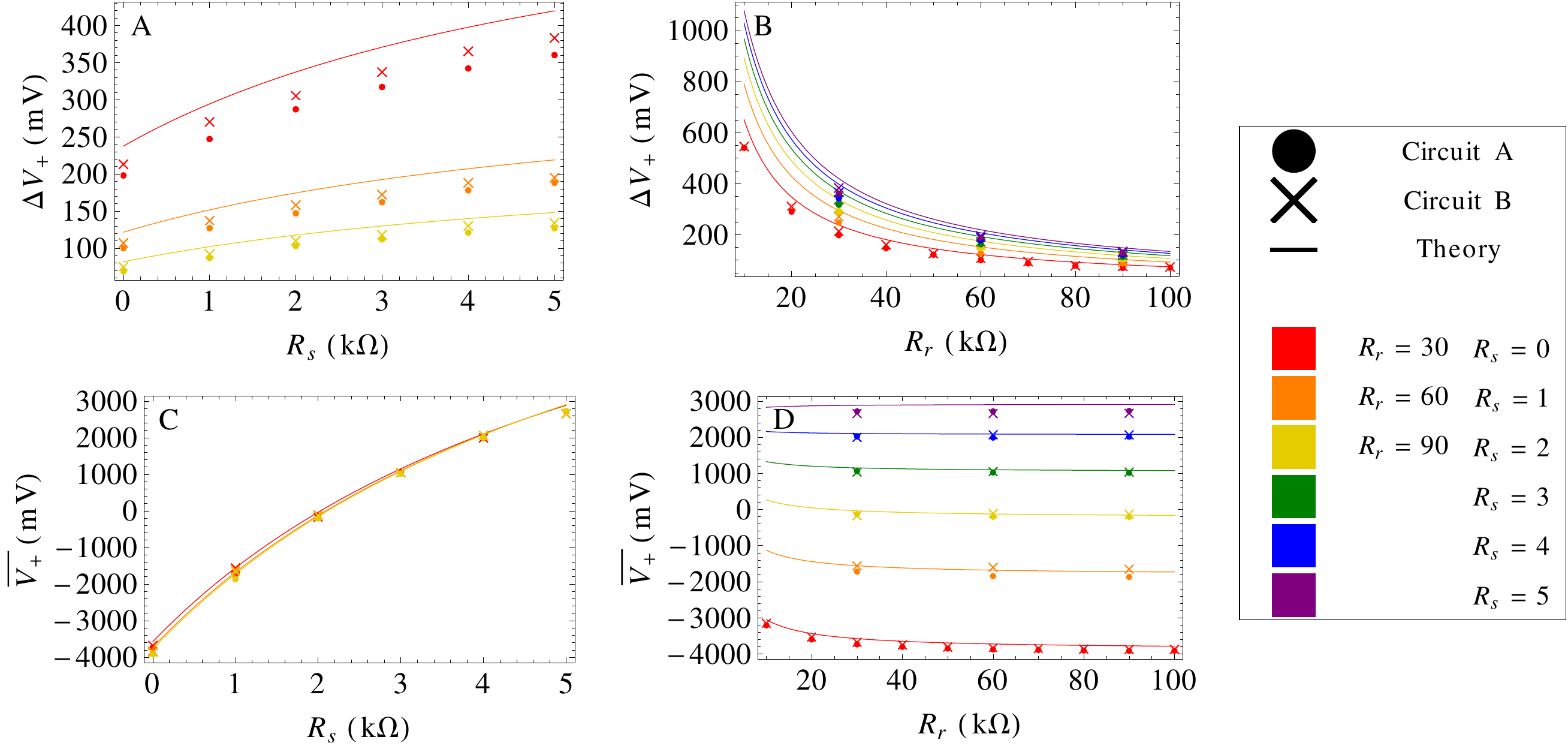}
 \caption[Schmitt trigger triggering windows]{Testing the two Schmitt trigger circuits: The colours indicate the trimpot resistances not shown on the axis in each graph.  $V_+$ is the voltage at the non-inverting input of the comparator.  Plots A and B show the range of the trigger window, and plots C and D show the set-point or average.}
 \label{fig:schmitt-plot}
\end{figure}

The schematic includes two \emph{trimpots} or variable resistors, $R_r$ and $R_s$, for tuning the triggering window. The circuit has been designed such that $R_r$ primarily effects the range of the triggering window, and $R_s$ primarily effects the set point or central value.  The effects on the trigger window of tuning trimpots were tested: a sinusoidal signal from a function generator was provided to the input of the Schmitt trigger, producing a square wave at $V_+$, the non-inverting input of the comparator.  The non-inverting input of the comparator is that which the input to the Schmitt trigger, on the inverting input of the comparator, is compared to.  The maximum, $V_H$, and minimum, $V_L$, of this square wave were recorded for various trimpot settings, and $\Delta V_+=V_H-V_L$ and $\overline{V_+}=\sfrac{\br{V_H+V_L}}{2}$ are plotted for both Schmitt trigger circuits in \autoref{fig:schmitt-plot}.

The smooth curves in the figure are derived from Kirchoff's circuit laws and  comparator behaviour \cite{horowitz} as we now show.  The capacitors are ignored in this derivation as they are active only at changing voltages, and act as open circuits in the maxima and minima of the square wave we are considering.  The output of the comparator is either high, $V_o=5V$, due to the $1\unit{k\Omega}$ pull up resistor on the output of the comparator, or low, $V_o=0V$.  $V_+$ can be seen as a weighted average of $12\unit{V}$, $-12\unit{V}$ and $V_o$, weighted by $\br{4.42\unit{k\Omega}}^{-1}$, $\br{R_s+2.26\unit{k\Omega}}^{-1}$, and $R_r^{-1}$ respectively.
\begin{equation}
 V_H=V_+\fbr{V_o=5\unit{V}}=\frac{12\unit{V}\br{4.42\unit{k\Omega}}^{-1} -12\unit{V}\br{R_s+2.26\unit{k\Omega}}^{-1}+5\unit{V}R_r^{-1}}{\br{4.42\unit{k\Omega}}^{-1}+\br{R_s+2.26\unit{k\Omega}}^{-1}+R_r^{-1}}
\end{equation}
and
\begin{equation}
 V_L=V_+\fbr{V_o=0\unit{V}}=\frac{12\unit{V}\br{4.42\unit{k\Omega}}^{-1} -12\unit{V}\br{R_s+2.26\unit{k\Omega}}^{-1}}{\br{4.42\unit{k\Omega}}^{-1}+\br{R_s+2.26\unit{k\Omega}}^{-1}+R_r^{-1}}.
\end{equation}

It can be seen from the slopes of the plots that as anticipated $R_r$ primarily affects the range, and $R_s$ primarily affects the set-point.  The differences between the two circuits, and between each and theory are fully expected and of no concern: the circuits were built using resistors whose values nominally matched those in the circuit diagram, but with large error bars.  As $R_r$ and $R_s$ would always be adjusted by observing $V_+$, this is not an inconvenience.

\section{Digital signal Processing}\label{sec:sigproc}
\begin{figure}[tb]
 \includegraphics[width=\columnwidth]{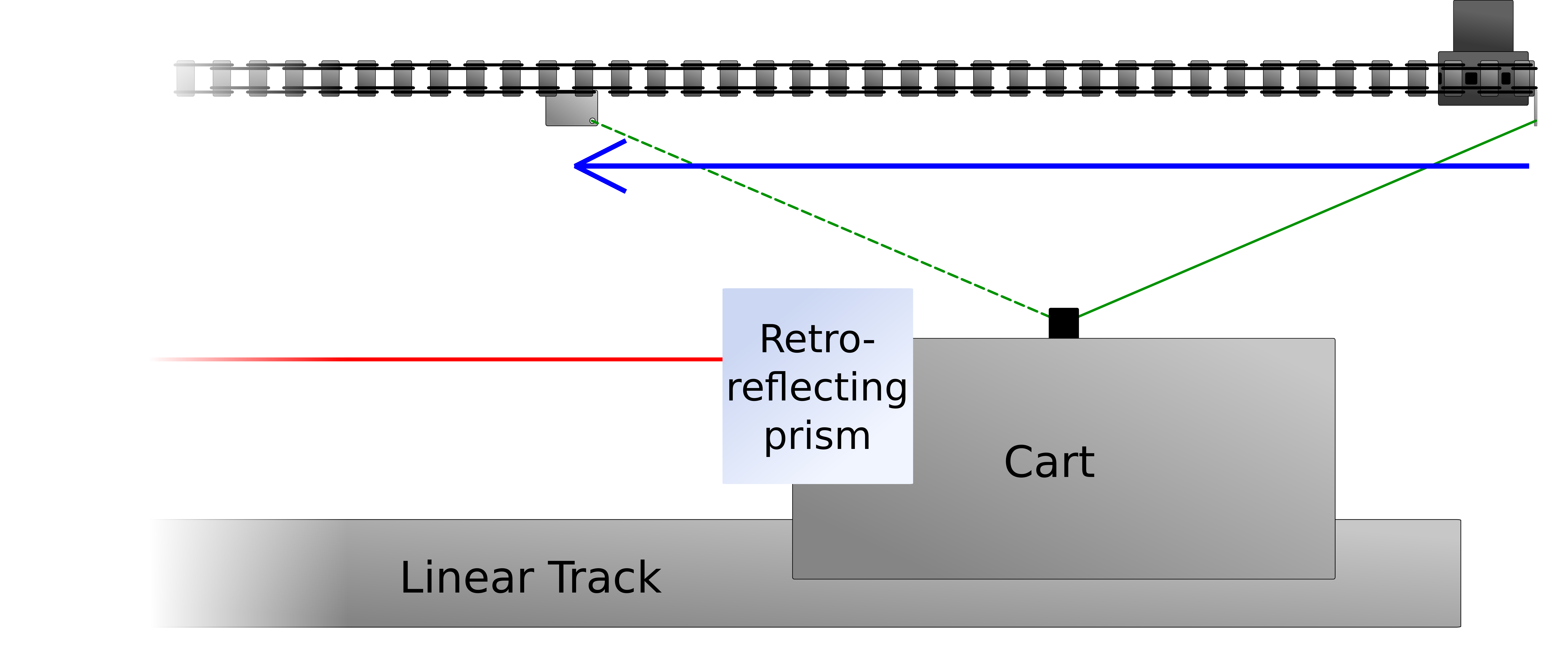}
 \caption[Previous wavemeter track system]{Under the previous linear track system the cart remained stationary throughout the time in which the section of chain to which the green wire was mounted moved through the distance indicated by the blue arrow.  It was also incredibly difficult to align laser beams such that the interference pattern was maintained throughout the entire path of the cart along the insufficiently-linear track.}
 \label{fig:track-turnaround}
\end{figure}

The signal processing board for the original wavemeter was able to be reused almost exactly as it was.  Under the previous linearly travelling cart system the signal processing board had to wait while the cart turned around and the fringes were unreliable due to vibrations in the system. This took some time as the cart was pulled by a chair-lift-like system as seen in \autoref{fig:track-turnaround},
where the cart remains stationary throughout the period in which the section of chain to which the green wire was mounted moved through the space indicated by the blue arrow.  This wait time was removed by shorting to ground a wire intended to come (under the cart system) from an optical switch.

A na\"ive interpretation of \autoref{eq:funknown} would suggest a frequency finding algorithm whereby fringes are counted for some predetermined amount of time, and then $N_\mathrm{unknown}$ and $\nu_\mathrm{HeNe}$ would be multiplied, and the result divided by $N_\mathrm{HeNe}$.  The division would not be performed first since the two fringe counts are expected to be similar -- both count fringes from red light in our experiments -- and this would require implementing floating point numbers which are not natively supported by the microcontroller.  Even performing the operations in the order specified would be time consuming since the numbers involved are sufficiently large that they span three 8-bit (1-byte) memory locations, and therefore require customized subroutines to perform even simple arithmetic.  Furthermore, division is not even implemented on the inexpensive microcontroller that we use.

\begin{figure}
 \centering
 \includegraphics[width=\columnwidth]{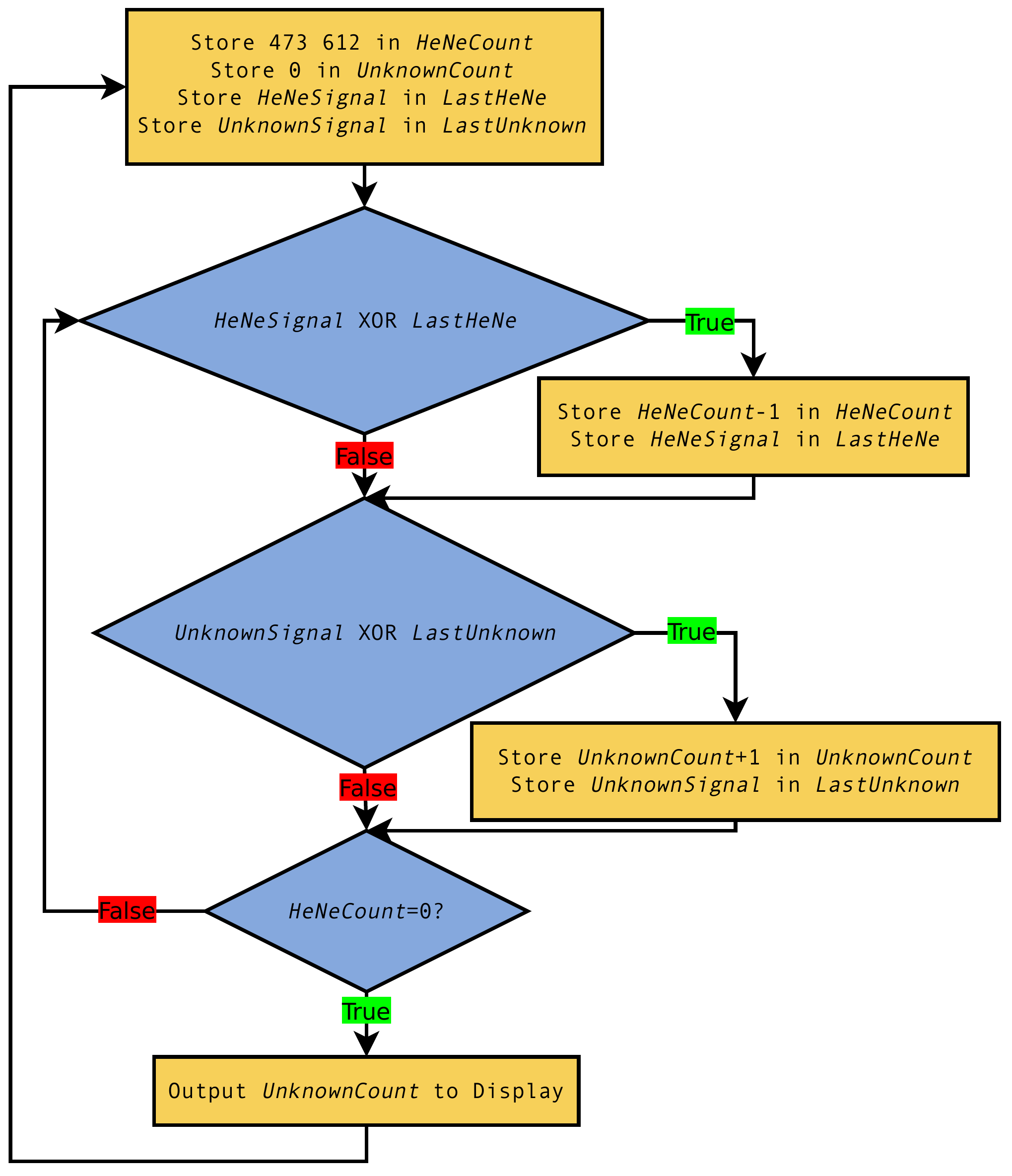}
 \caption[Flowchart of the microcontroller algorithm]{A simplified flowchart of the algorithm loaded onto the programmable microcontroller.  The frequency of the reference HeNe laser is $473\,612\unit{GHz}$ \cite{fox}.}
 \label{fig:microcontrollerFlowchart}
\end{figure}

To solve this, we use the increment/decrement scheme suggested by Fox \emph{et al.} \cite{fox}.  Rewriting \autoref{eq:funknown} as
\begin{equation}
\nu_\mathrm{unknown}=\frac{\nu_\mathrm{HeNe}}{N_\mathrm{HeNe}}N_\mathrm{unknown},
\end{equation}
makes it clear that it is far simpler to have the algorithm count fringes until it has counted a number of fringes from the \ce{HeNe} laser equal to the frequency of the \ce{HeNe} laser in some units; we choose gigahertz.  The division $\sfrac{\nu_\mathrm{HeNe}}{N_\mathrm{HeNe}}$ then gives $1\unit{GHz}$, and $\nu_\mathrm{unknown}$ is then simply $N_\mathrm{unknown}\unit{GHz}$.  This is called an increment/decrement scheme because it is often easier in computer science to test if a number is equal to zero than equal to some other non-zero number.  This means that the scheme is best implemented by starting a counter for the reference HeNe laser at $473\,612$, the frequency of the laser in gigahertz \cite{fox}.  This counter is then decremented while a second counter for the unknown laser is incremented.  When the first counter reaches zero, all counting stops.  A simplified flowchart of the algorithm loaded onto the programmable microcontroller, is shown in \autoref{fig:microcontrollerFlowchart}.

We drive the rocker at a frequency and amplitude such that maximum rate of change in path difference is comparable to that used by the previous track and cart system. This ensures that the frequency of the fringes is sufficiently slow that the same chip as used in the previous track and cart system (operating at the same clock speed) is capable of detecting all pulses from the Schmitt triggers.  This frequency happens to be slightly slower than the resonant frequency of the rocker, which introduced some bounce to the turn-around of the rocker.  This was undesirable as the bouncing added noise to the laser fringes.  To correct for this, masses were added to the armatures of the rocker, reducing its resonant frequency to be sufficiently close to that of the driving frequency that bouncing ceased.

\section{Saturated absorption spectroscopy}\label{sec:satspec}

\begin{figure}
 \begin{center}
  \includegraphics[width=0.4\columnwidth]{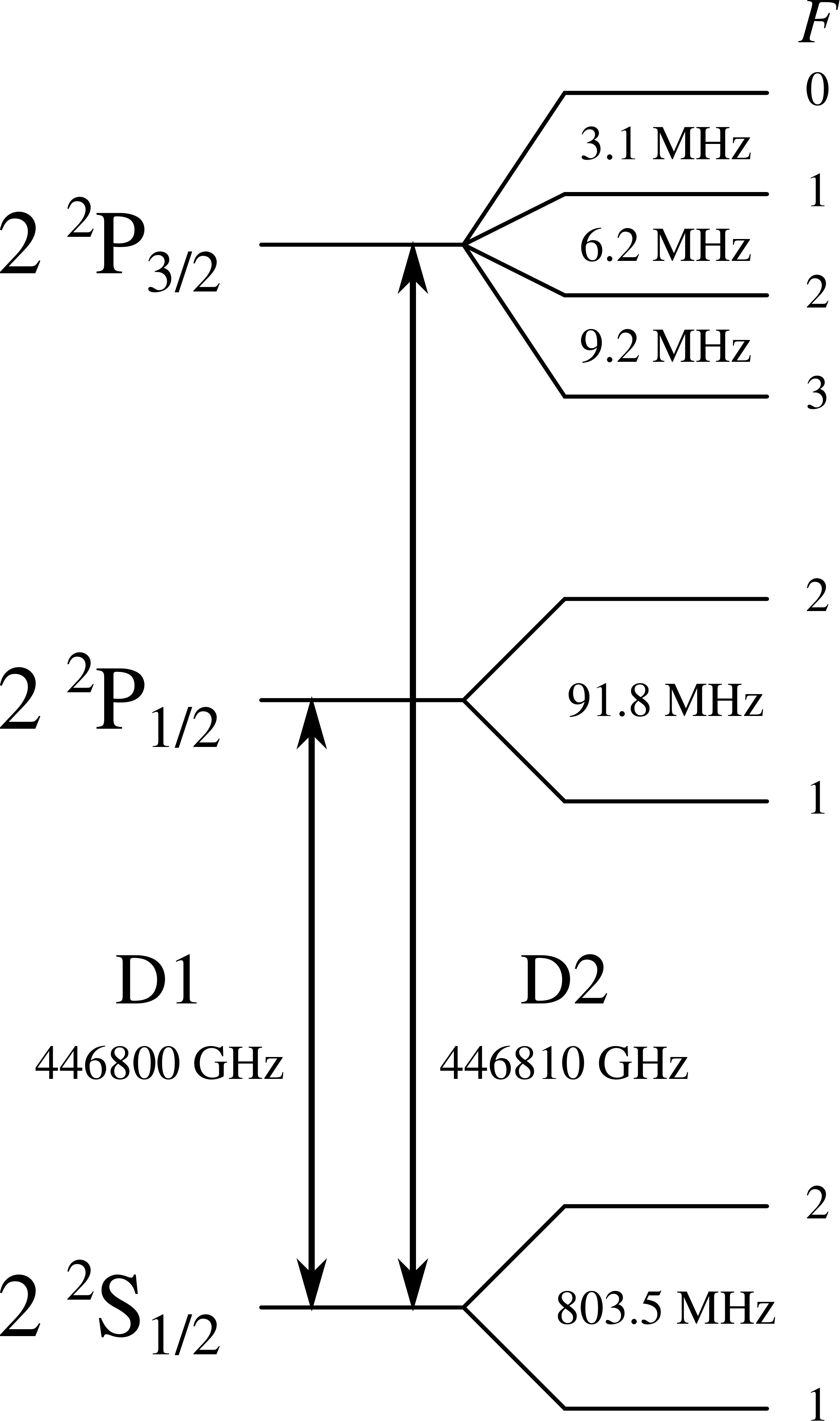}
 \end{center}
 \caption[Energy level diagrams of \ce{^7li}]{A subset of the energy levels of \ce{^7Li} showing the D1 and D2 lines.}
 \label{fig:lithium-levels}
\end{figure}

In order to test the accuracy of the wavemeter, we used saturated absorption spectroscopy \cite{demtroder} of the D1 and D2 lines of lithium since this will be required for locking the laser to the D2 line in the cold collision experiment.  \autoref{fig:lithium-levels}
shows these lines in \ce{^7Li}, the more abundant (92.4\%) isotope.  The hyperfine splittings may be found in Ref.~\citenum{libbrecht}.

\begin{figure}
 \includegraphics[width=\columnwidth]{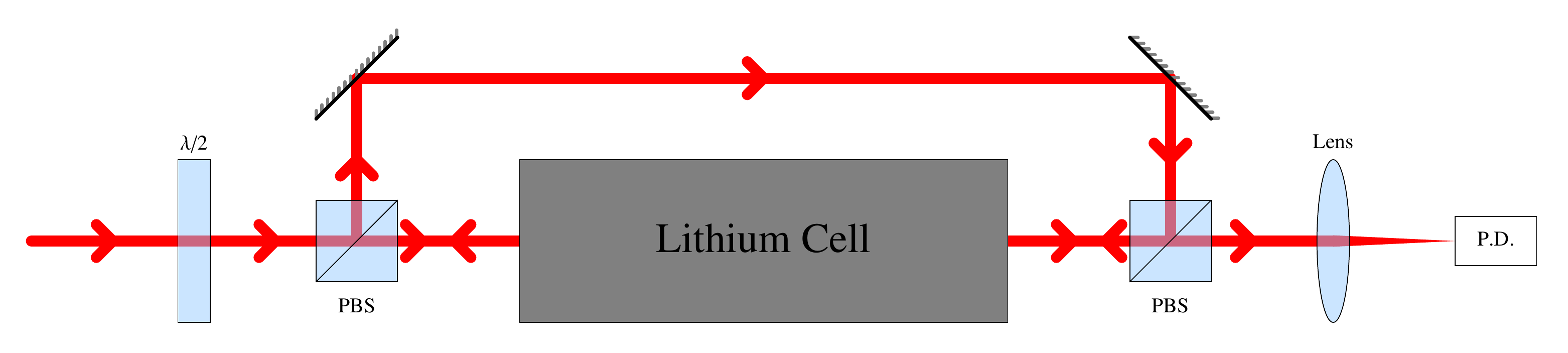}
 \caption[Saturation spectroscopy setup]{The optical setup for saturated absorption spectroscopy in which a half-wave plate, $\sfrac{\lambda}{2}$, is used to adjust horizontal and vertical polarizations before those are split by a polarizing beam splitter, PBS.  A second PBS directs the vertically polarized pump beam back through the Lithium cell along the same path as the much weaker counter-propagating horizontally polarized probe beam which passes through the second PBS to be focused through a lens into the photodiode.}
 \label{fig:satspec-setup}
\end{figure}

Saturated absorption spectroscopy of lithium was performed with counter-pro\-pa\-ga\-ting pump and probe beams through a lithium cell as shown in \autoref{fig:satspec-setup},
with the lithium cell heated to and maintained at $590\unit{K}$ to vaporize the lithium inside.  The external cavity diode laser is scanned using a function generator going to both the stack and current of the laser.  The piezoelectric stack controls the length of the external cavity, and adjusting the two together allows for a greater scanning range while remaining in single mode operation.  Scanning is performed over a couple of gigahertz around the transition (D1 or D2) frequency.  When the pump beam is detected by a photodiode (not shown in the figure), a Doppler-broadened absorption dip can be seen as the laser is scanned, with a full-width half-maximum of \cite{hollas-doppler}
\begin{equation}
\Delta \nu = \sqrt{\frac{8k_B T ln\fbr{2}}{m}}\frac{\nu_0}{c} \approx 3\unit{GHz},
\end{equation}
where $k_B$ is the Boltzmann constant, $T$ is temperature of the lithium cell, $m$ is the mass of lithium, and $\nu_0$ is the (rest frame) frequency of the transition.  The beam addresses at each laser frequency within the dip those atoms with axial velocity such that the Doppler shifted laser frequency matches a transition frequency.  This feature is too broad to determine the accuracy of the wavemeter, and far too broad for locking to in cold collision experiments given that transitions are required in \emph{cold} lithium which experiences minimal Doppler broadening.

When addressing the same transition, the counter-propagating probe beam addresses atoms with an axial velocity which is the negative of that of the atoms addressed by the pump beam.  Therefore when the incident lab-frame laser frequency is within the natural linewidth of a transition, the pump and probe beams address the same atoms which have no Doppler broadening.  The far stronger pump beam is absorbed while depleting and optically pumping the transition, so that a transmission \emph{peak} is visible in the probe beam.

The natural linewidth of the D1 and D2 transitions is $5.8\unit{MHz}$ \cite{wijngaarden2009}, so the closely spaced \ce{2\,^2P_{$\sfrac{3}{2}$}} hyperfine levels in the D2 line cannot be properly resolved, but (\ce{2\,^2S_{$\sfrac{1}{2}$}}, $F=1$) \ce{<->} \ce{2\,^2P_{$\sfrac{3}{2}$}} and (\ce{2\,^2S_{$\sfrac{1}{2}$}}, $F=2$) \ce{<->} \ce{2\,^2P_{$\sfrac{3}{2}$}} can be.  In D1, all four hyperfine transitions can be resolved.

Observing the probe beam also permits \emph{cross-over dips} to be seen whenever the laser frequency is halfway between two transitions with different hyperfine ground states.  When this is the case, there are some atoms with an axial velocity that allow them to Doppler shift in opposite directions for the two counter-propagating beams such that both beams address them but on the two different transitions.  The pump beam is then absorbed in exciting them to a state from which they can decay into either hyperfine ground state.  As the pump beam continues to excite those that fall into the ground state that it is addressing, atoms are preferentially pumped over to the ground state addressed by the probe beam, which therefore has increased absorption, resulting in a dip at the photodiode.

The photodiode signal of the probe beam therefore shows the same overall Doppler broadened dip as the pump beam, but additionally shows some Doppler-free structure.

\section{Results}\label{sec:waveresults}

\begin{figure}[t]
 \includegraphics[width=\columnwidth]{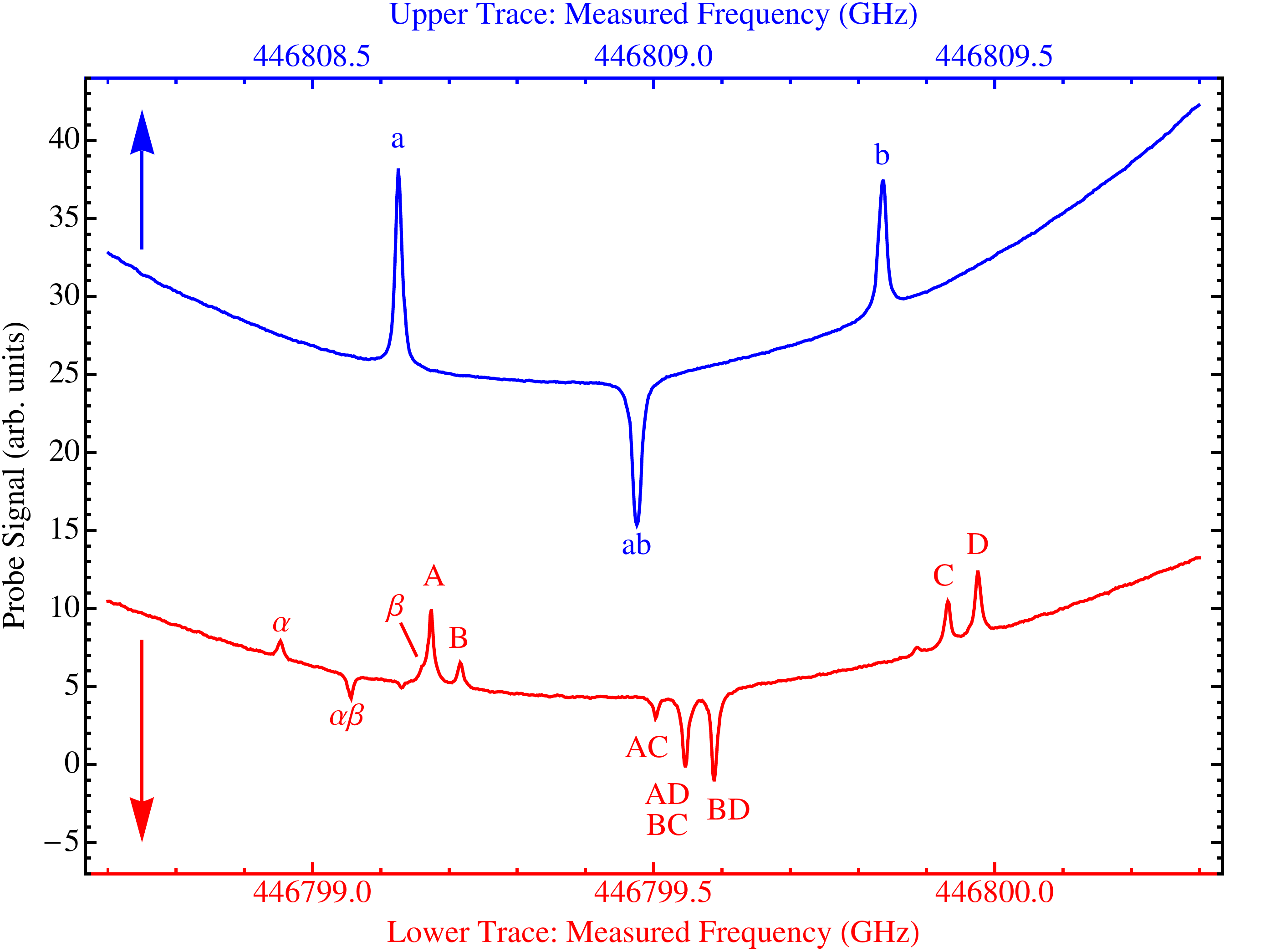}
 \caption[Testing the redesigned wavemeter with saturated absorption spectroscopy]{Validating the redesigned wavemeter using saturation spectra of the lithium D1 and D2 lines. Features in D2 (D1) spectra are marked in lower (upper) case.  Features of \ce{^7Li} (\ce{^6Li}) are marked in Roman (Greek) letters.  The central value of frequency plotted is the midpoint of the range of several wavemeter readings, and the plotted frequency range for each trace was determined by observing the motion of the peak on an OSA with a known FSR of $7.5\unit{GHz}$.}
 \label{fig:wavemeter-satspec}
\end{figure}

Saturation spectra were taken in the neighbourhood of $446\,800\unit{GHz}$ and $446\,810\unit{GHz}$ according to the redesigned wavemeter.  The resulting traces were then saved via an oscilloscope which also recorded, for the sake of converting the horizontal axis of time into laser frequency, the function generator sawtooth output used to control the laser frequency.  The midpoint between maximum and minimum of the sawtooth was assigned a frequency using the wavemeter with the function generator turned off, taking the midpoint of the range over several readings.  Operating under the assumption that over small ranges the laser frequency varies linearly with the function generator output, the range between maximum and minimum of the sawtooth was assigned a frequency range based on the travel of the peak from an optical spectrum analyser (OSA) with a free spectral range (FSR) of $7.5\unit{GHz}$ as observed on a oscilloscope while the function generator was on.  This resulted in the two traces shown in \autoref{fig:wavemeter-satspec}.
It can be seen by the central locations of the cross-over dips relative to the peaks that the assumption of a linear relationship between function generator output and laser frequency is a reasonable one.  The discrepency between the hyperfine splittings given in \autoref{fig:lithium-levels} and those suggested by \autoref{fig:wavemeter-satspec} is due to an error in reading the range of frequency values from the OSA of about 15\%.  The values obtained from the wavemeter are the centrally plotted $446\,809\unit{GHz}$ and $446\,799.5\unit{GHz}$.  The upper trace shows the two expected peaks of the D2 spectrum: \begin{align*}
 a)\qquad&(\cee{2\,^2S_{\sfrac{1}{2}}}, F=1) \longleftrightarrow \cee{2\,^2P_{\sfrac{3}{2}}}\\
 \textrm{and }b)\qquad&(\cee{2\,^2S_{\sfrac{1}{2}}}, F=2) \longleftrightarrow \cee{2\,^2P_{\sfrac{3}{2}}},
\end{align*}
as well as their associated cross-over dip, $ab$.  The lower trace shows the expected peaks of the D1 spectrum for
\begin{align*}
 A)\qquad&(\cee{2\,^2S_{\sfrac{1}{2}}}, F=2) \longleftrightarrow (\cee{2\,^2P_{\sfrac{1}{2}}}, F=1),\\
 B)\qquad&(\cee{2\,^2S_{\sfrac{1}{2}}}, F=2) \longleftrightarrow (\cee{2\,^2P_{\sfrac{1}{2}}}, F=2),\\
 C)\qquad&(\cee{2\,^2S_{\sfrac{1}{2}}}, F=1) \longleftrightarrow (\cee{2\,^2P_{\sfrac{1}{2}}}, F=1),\\
 \textrm{and }D)\qquad&(\cee{2\,^2S_{\sfrac{1}{2}}}, F=1) \longleftrightarrow (\cee{2\,^2P_{\sfrac{1}{2}}}, F=2).
\end{align*}

The crossover dip $AD$ occurs at the same frequency as dip $BC$, so only three distinct dips are observed in the D1 spectrum.  In the same lower trace as the D1 spectrum, a much smaller D2 spectrum for \ce{^6Li} ($\alpha$, $\beta$, $\alpha\beta$) can be observed with its own central frequency and \ce{2\,^2S_{$\sfrac{1}{2}$}} $F=1$, $F=2$ spacing, where the upper transition, $\beta$, is mostly subsumed in the $A$ peak.

One drawback to the use of a rocker is the occurrence of frequent turnarounds which, while they do not introduce delay as in the previous linear track design, do introduce some error in the calculated frequency.  At each turnaround the fringe pattern also reverses with an essentially random phase, and the microcontroller has only the option of counting or not counting a fringe.  This allows for the introduction of up to $\pm0.5\unit{GHz}$ of error at each turnaround, which over many repetitions would average to zero.  We model this as introducing an error of $\pm0.5\unit{GHz}$ with equal probability of positive or negative value at each of the 15 turnarounds for each result.  This can be seen to be a shifted binomial distribution with number of tests, $n=15$, and probability of success, $p=\frac{1}{2}$.  The root-mean-square error is therefore $\sqrt{np\br{1-p}}\unit{GHz}=\sqrt{\sfrac{15}{4}}\unit{GHz}\approx 1.9\unit{GHz}$.  This is the primary source of error for the wavemeter.

\section{Conclusions}

While we had at first hoped to be able to make some slight modifications to the previous wavemeter to make it easier to use, a major redesign in which the optical train was completely replaced and the electronics improved was found necessary.

The new wavemeter is considerably easier to align and performs well. It has been tested for accuracy by using saturated absorption spectroscopy with an external cavity diode laser as it was tuned across the D1 and D2 lines of \ce{^7Li}.

\chapter{Conclusions}\label{ch:conclusions}\enlargethispage{\baselineskip}
A long term goal of the lab, that of constructing and using a lithium magneto-optical trap for cold collision studies, has been advanced on two fronts.  In \autoref{part:rydberg} of the thesis we described an effective method for calculating bound-to-continuum cross-sections for charged binary systems by placing the system in an infinite potential well, and examining transitions to states above the binding energy that in this model became bound, which aided in performing numerical calculations.  This approach was first verified for photo-ionization of a hydrogen atom, and then applied to the case of photo-dissociation a heavy Rydberg system, \ce{Li+...I-}, which are, to the best of our knowledge, the first heavy Rydberg photo-dissociation cross-sections presented.  The oscillatory nature of the cross-sections at both initial states considered suggests an interesting possibility of selective control using light with an energy which cannot dissociate the ion pair by virtue of targeting an upwards transition to a cross-section minimum.  This light would therefore only force a particular desired downward transition.  These calculations lay excellent groundwork for minor adjustments to be made to the parameters of the program found in \autoref{app:code} in order to calculate cross-sections for \ce{Li+...Li-}, or indeed any other ion pair.

The second area in which progress has been made towards cold collision studies is that of the redesign and construction of a wavemeter.  The wavemeter previously built in the lab took far too long (on the order of half a day) to align by virtue of the fact that it was impossible to properly align it for all positions of the insufficiently straight travelling Michelson interferometer it employed, requiring a compromised minimal misalignment at all positions.  The redesign made use of a rocker system recovered from a BOMEM FT-IR spectrometer to replace the optical layout and hardware of the previous wavemeter. This resulted in increased reliability and ease of use: the new wavemeter can be aligned in mere minutes, and a problem with cross-talk between two Schmitt triggers has been resolved.  The new wavemeter has been tested through saturated absorption spectroscopy of lithium, which was the chosen technique because it not only verifies the wavemeter, but also constitutes a key part of the next step of the experiment, that of locking one of the diode lasers directly to a feature within the saturated absorption spectrum.

{\backmatter
\cleardoublepage
\phantomsection
\addcontentsline{toc}{chapter}{\bibname}
\bibliography{refs}{}
\bibliographystyle{thesis}}

\begin{appendices}
\chapter{Mathematica Code}\label{app:code}
\fancyhead[LO,RE]{\itshape\autoref{app:code}: Mathematica Code}
\vspace{-5pt}
\includegraphics[scale=0.9]{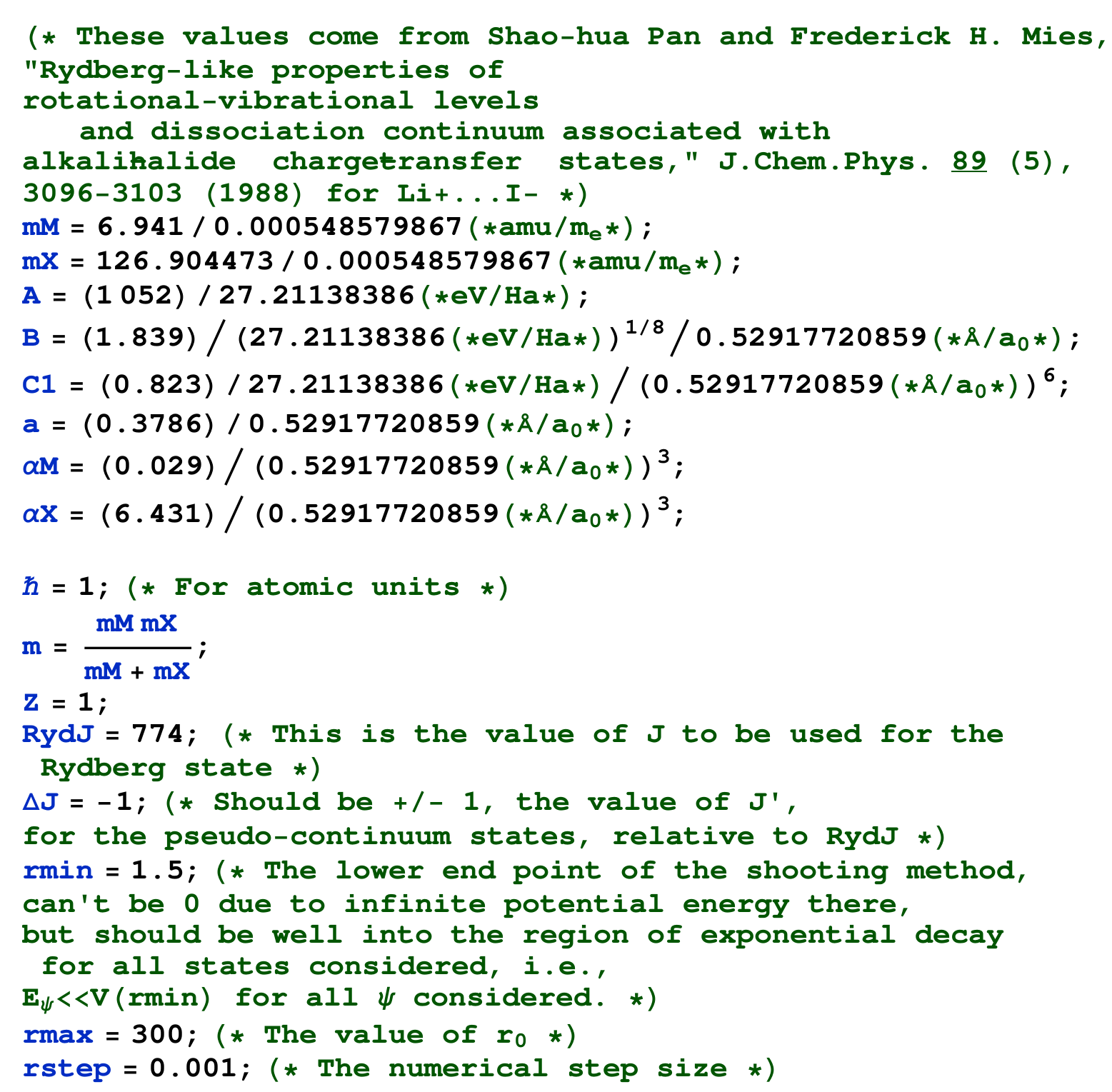}

\includegraphics[scale=0.85]{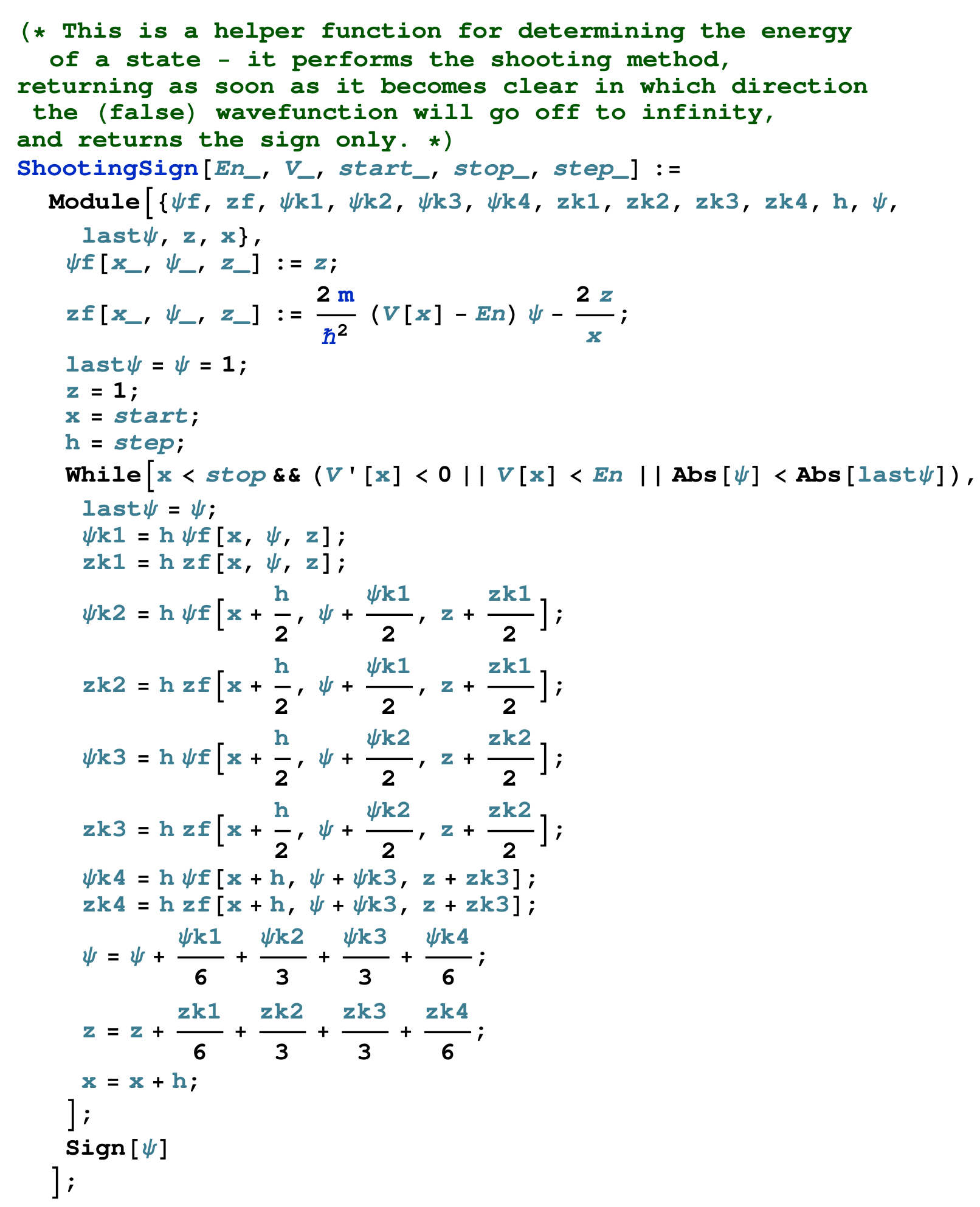}

\includegraphics[scale=0.85]{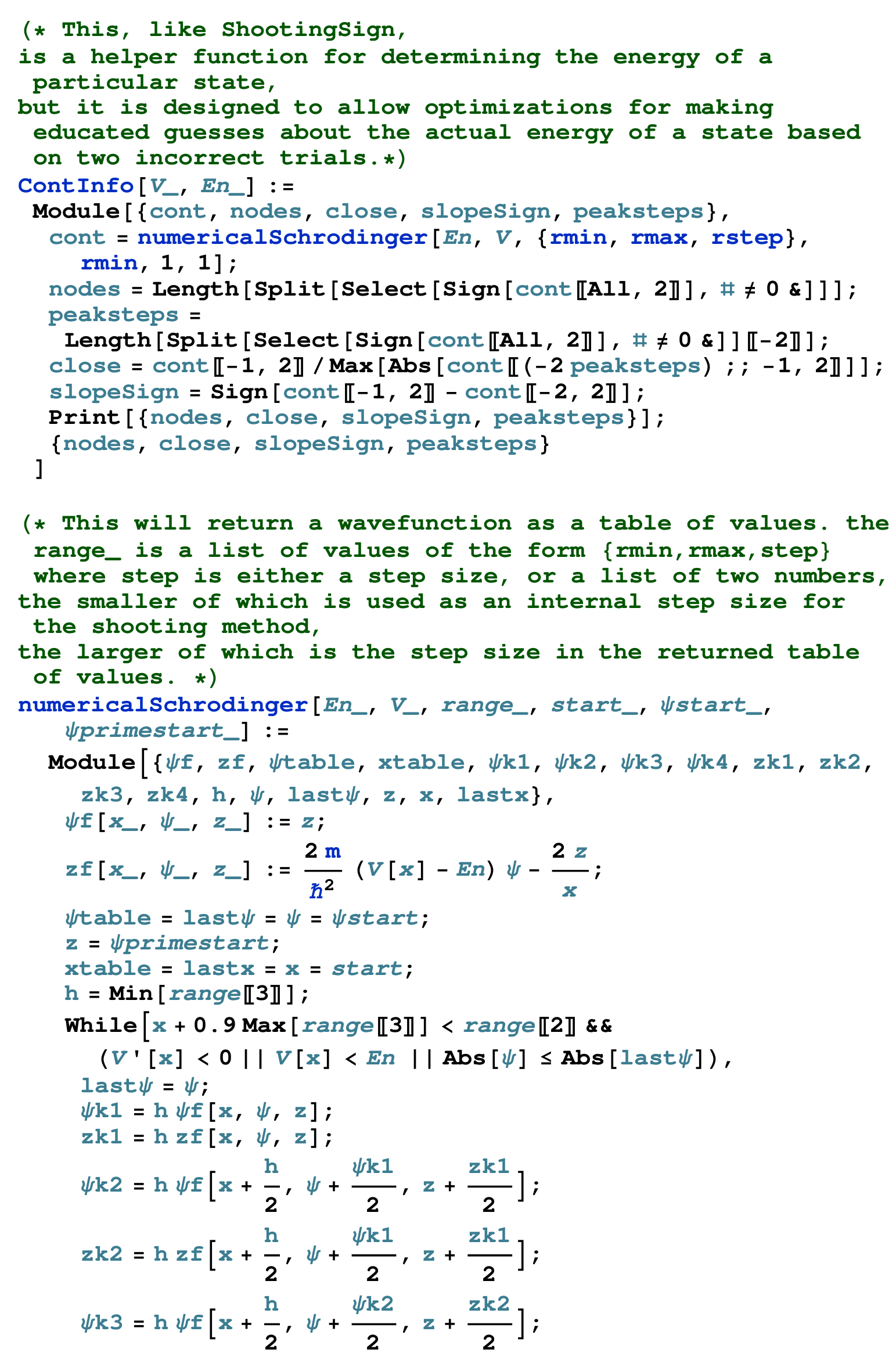}

\includegraphics[scale=0.85]{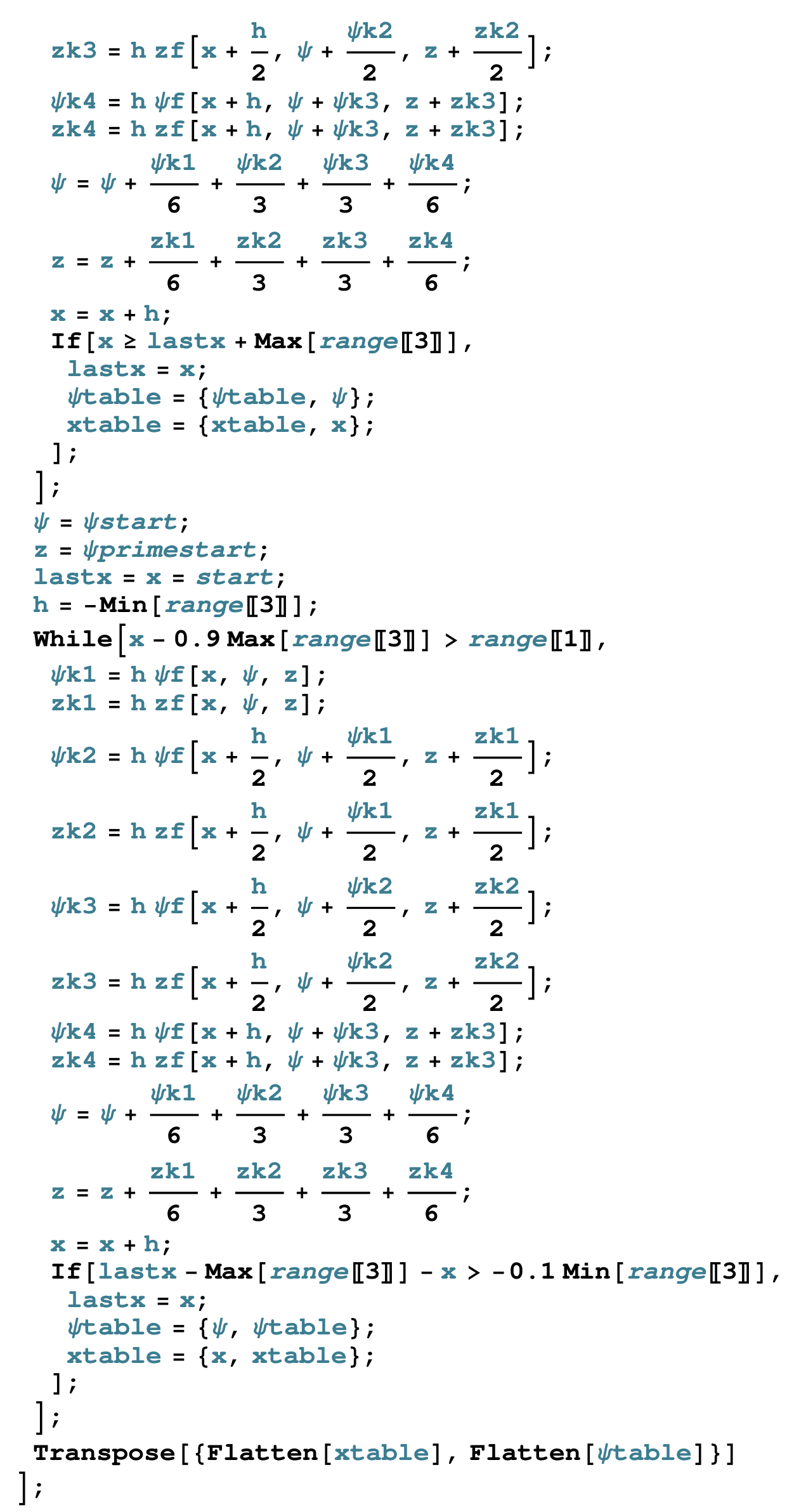}

\includegraphics[scale=0.85]{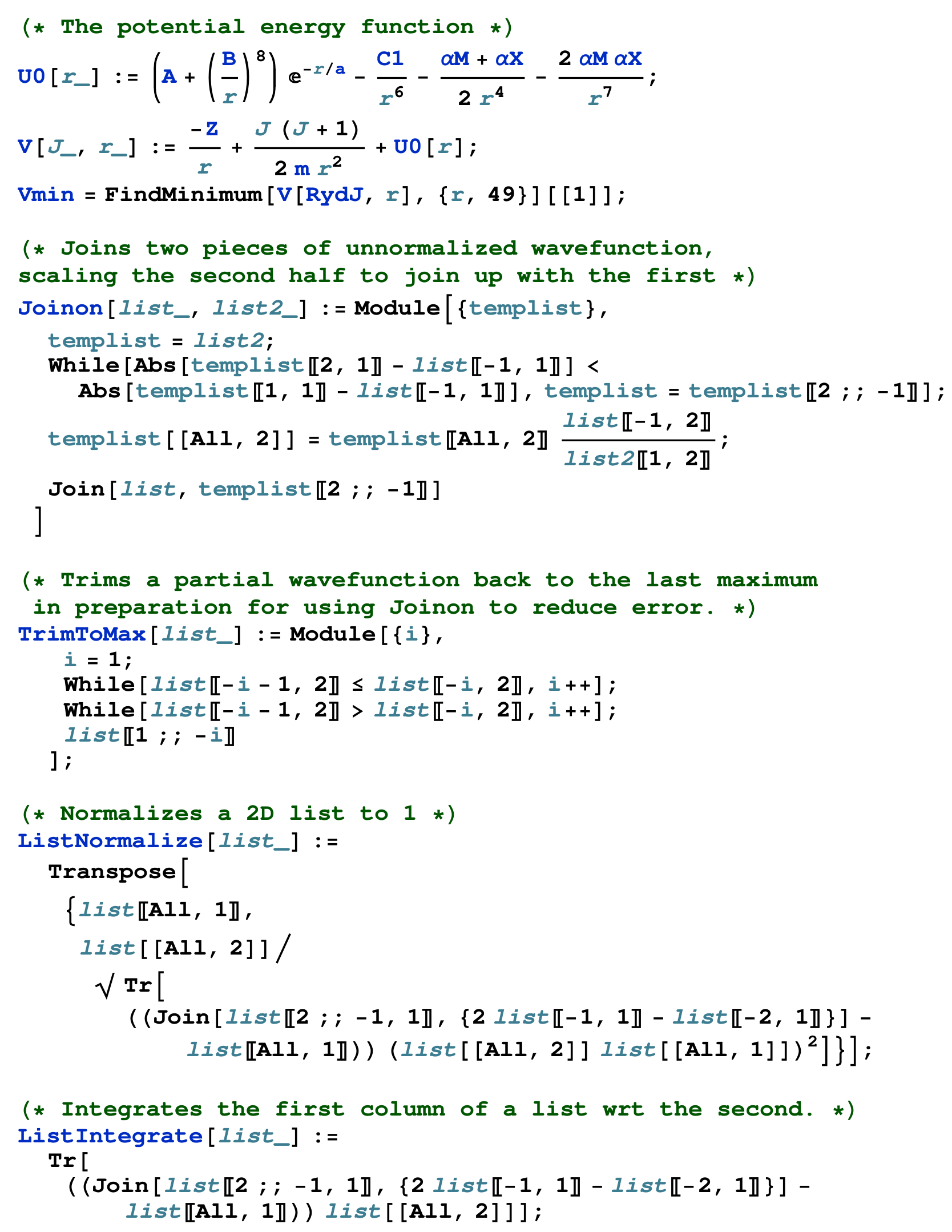}

\includegraphics[scale=0.85]{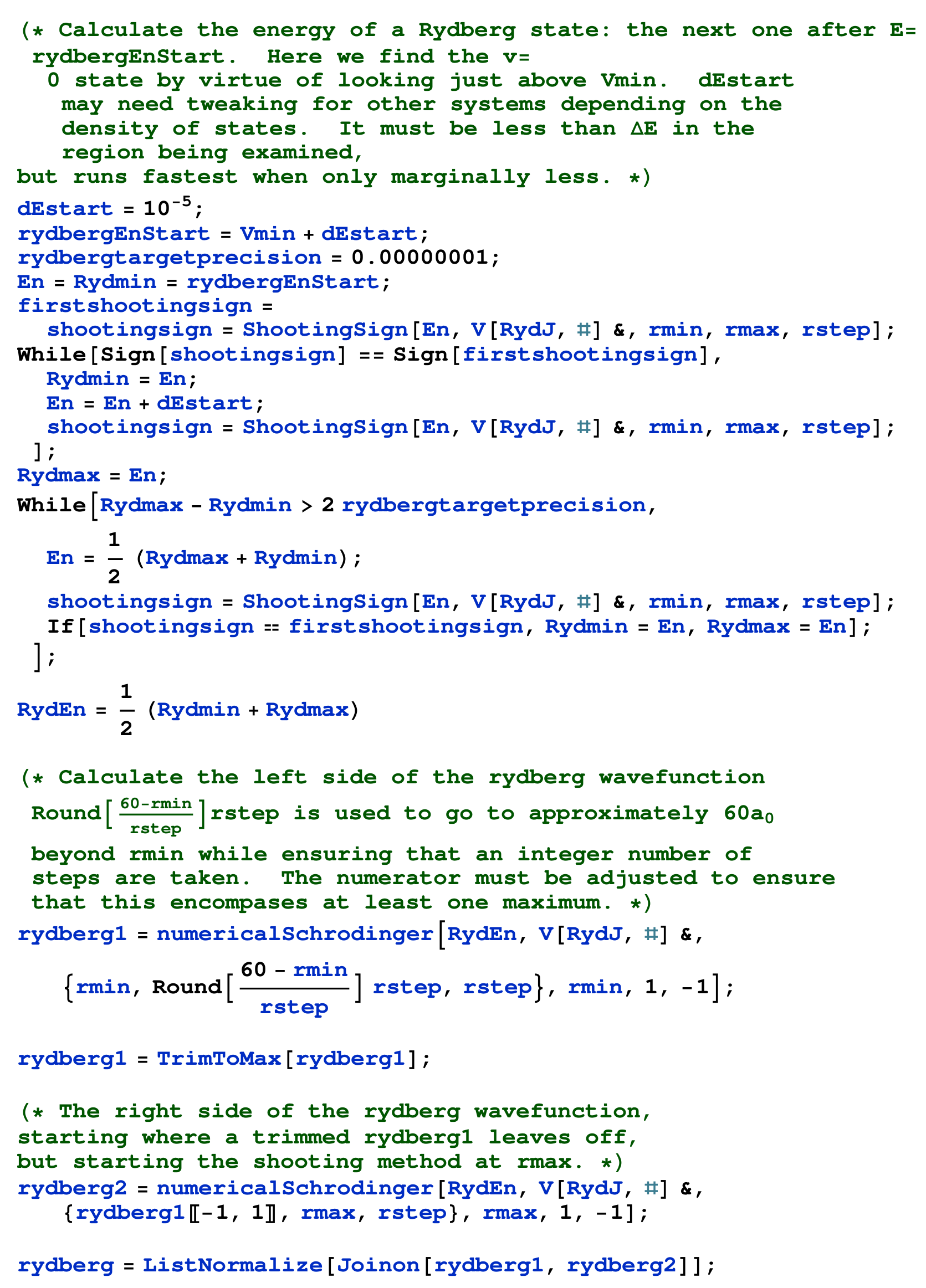}

\includegraphics[scale=0.85]{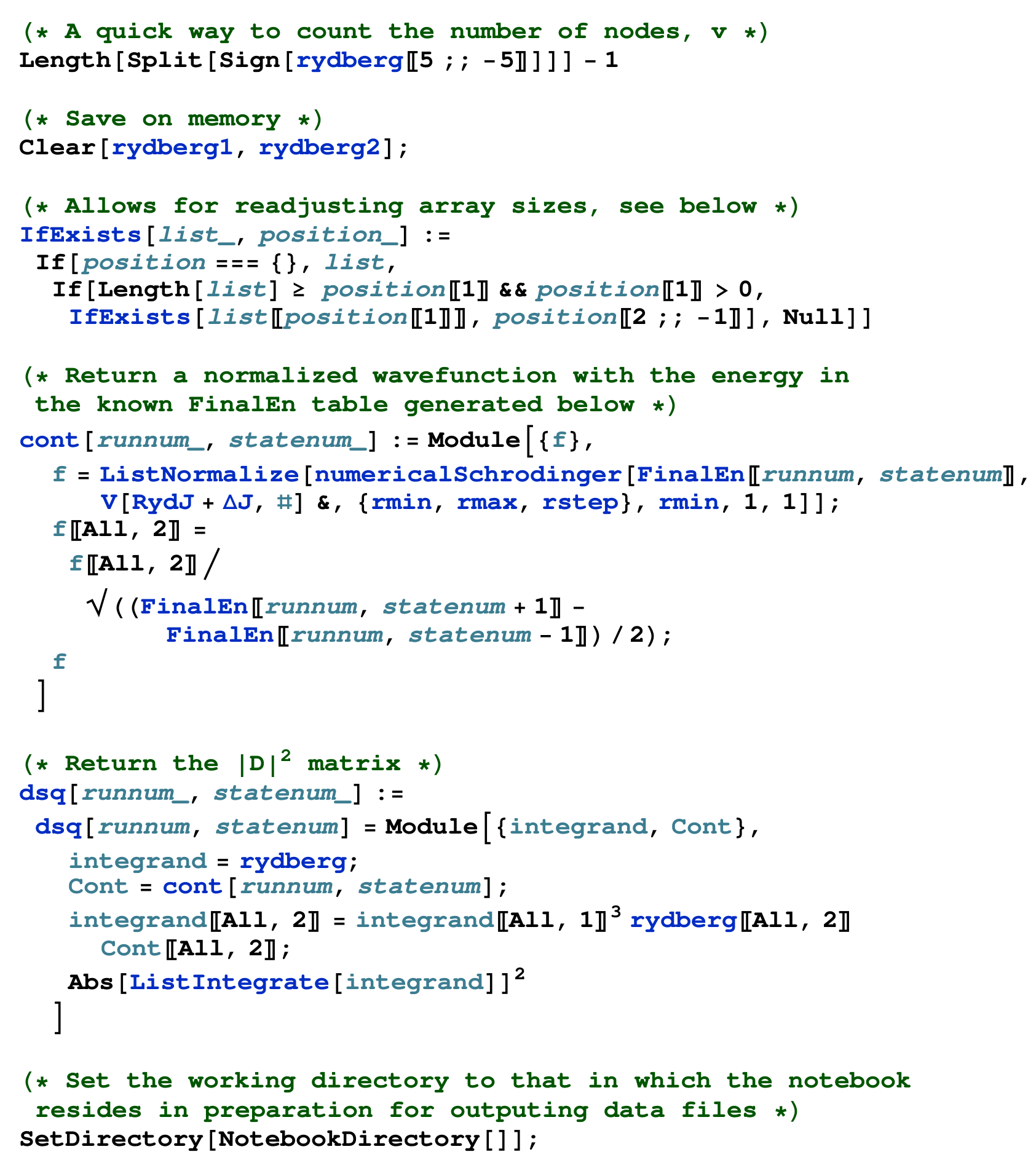}

\includegraphics[scale=0.85]{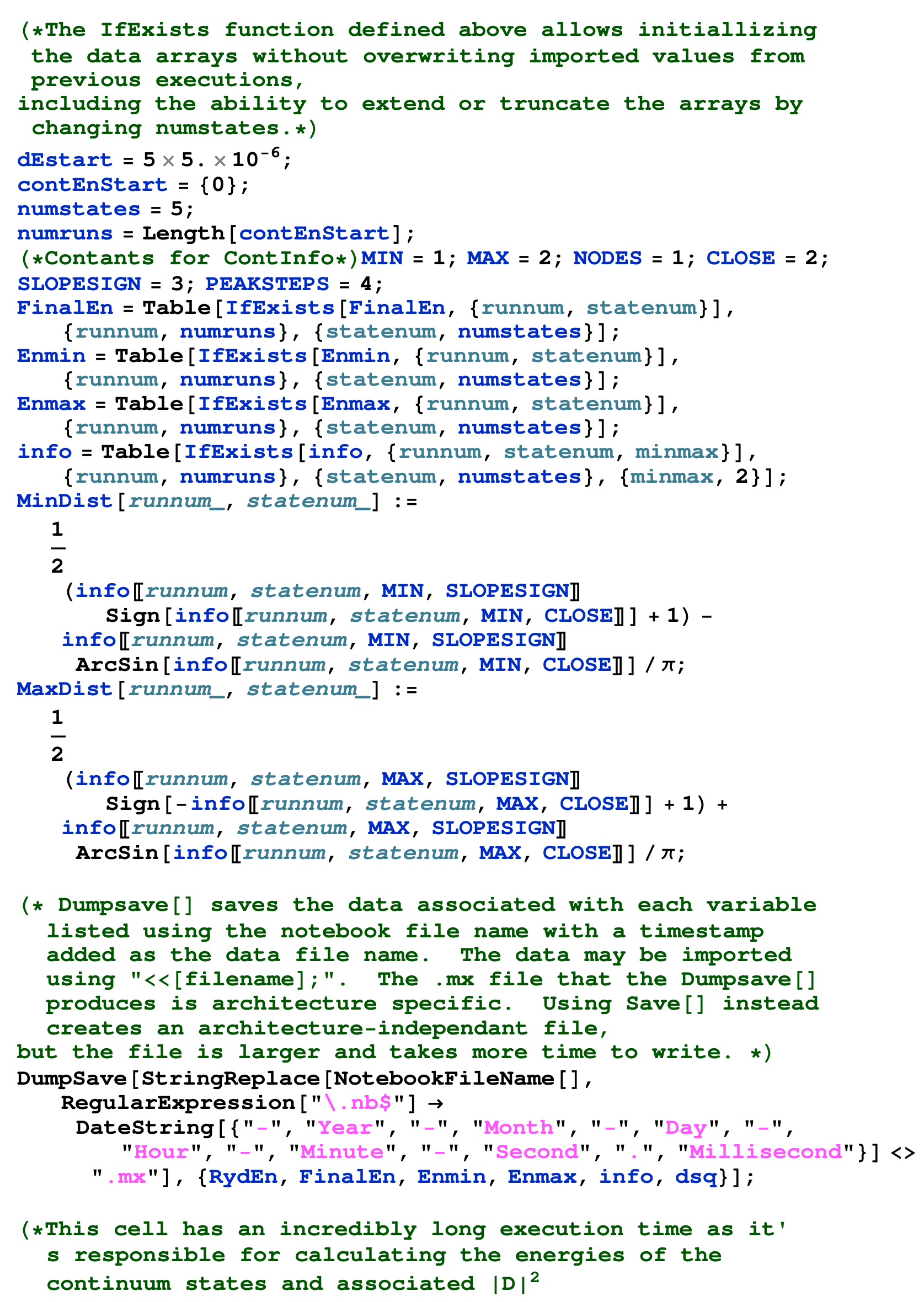}

\includegraphics[scale=0.85]{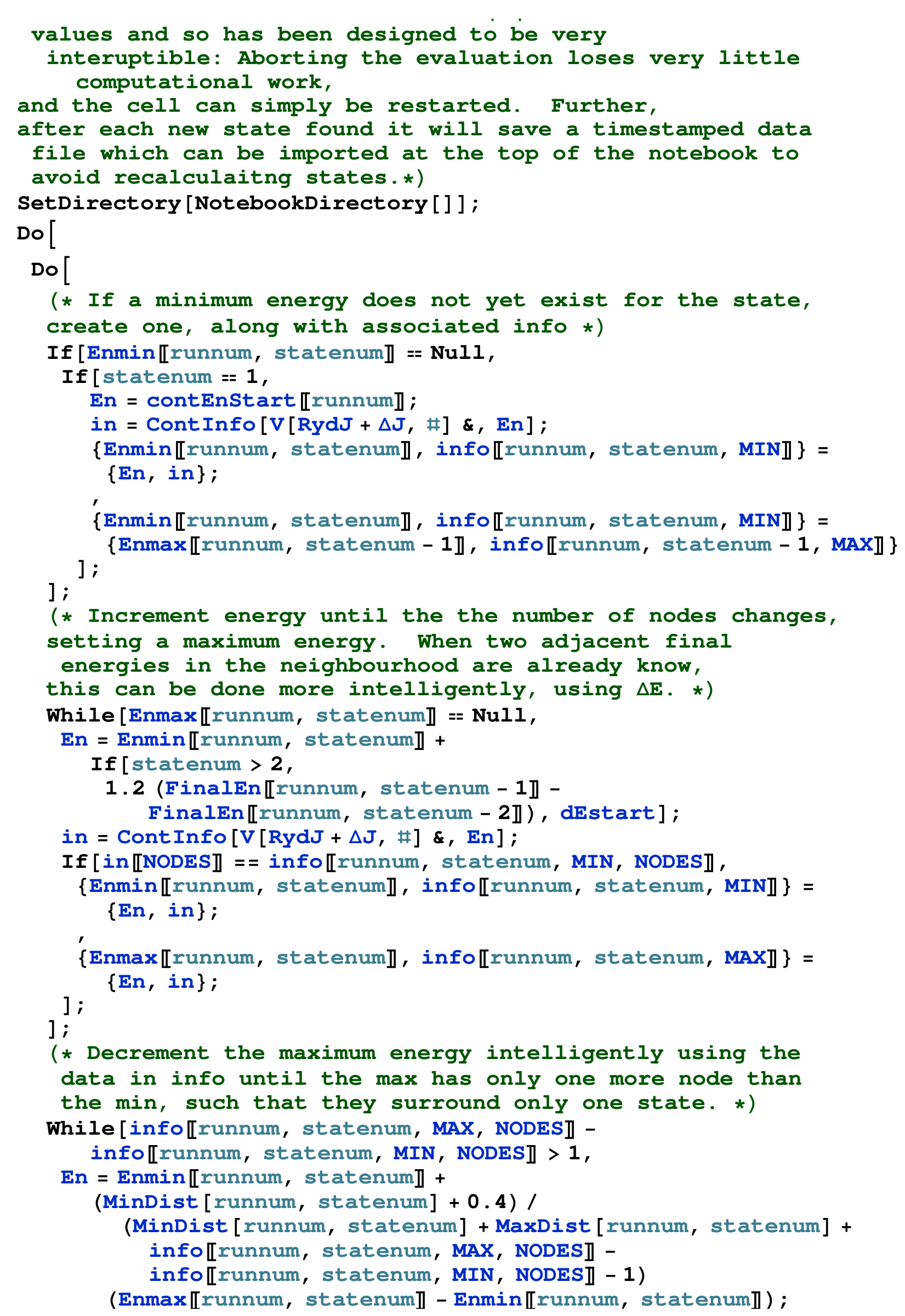}

\includegraphics[scale=0.85]{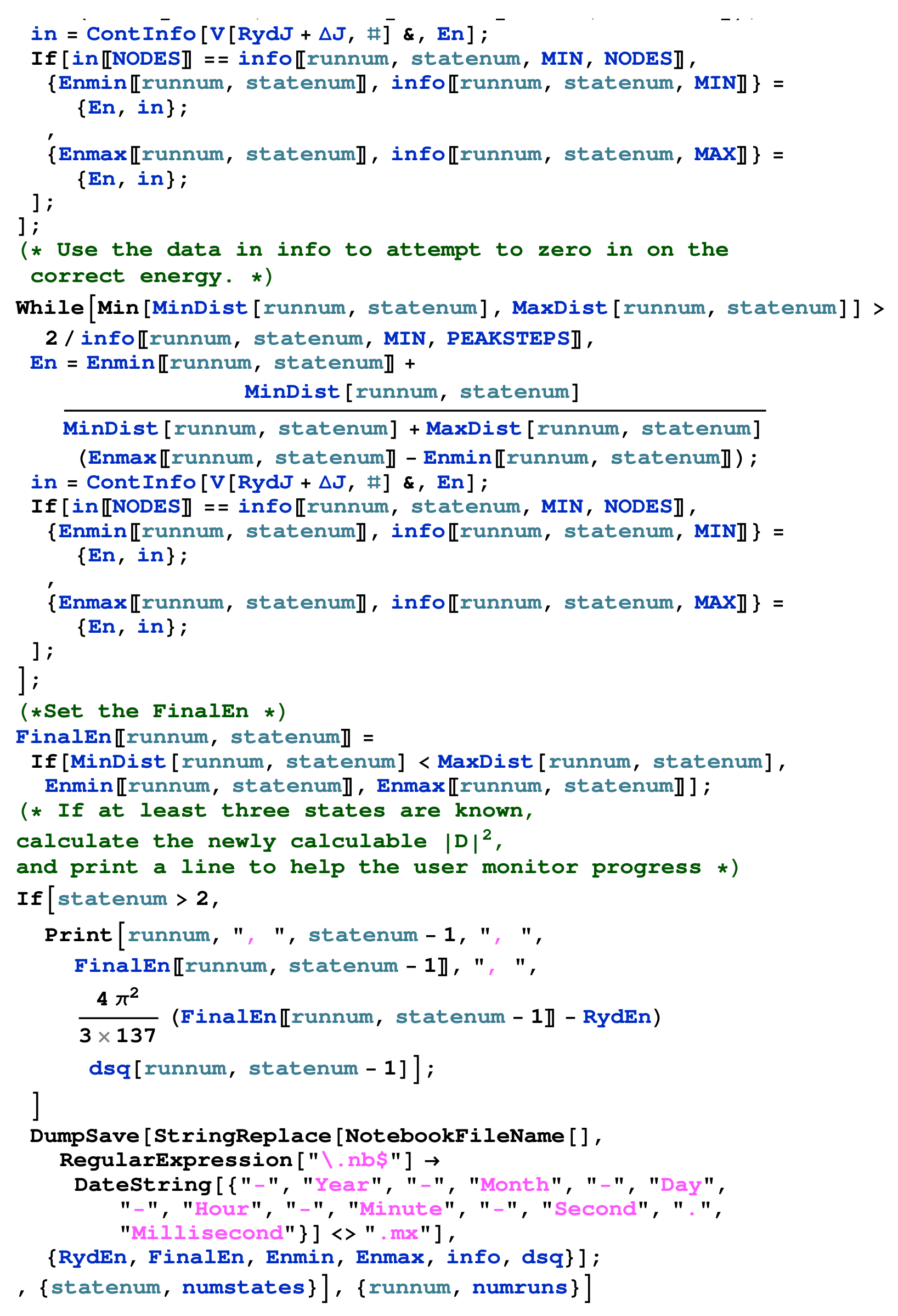}

\includegraphics[scale=0.85]{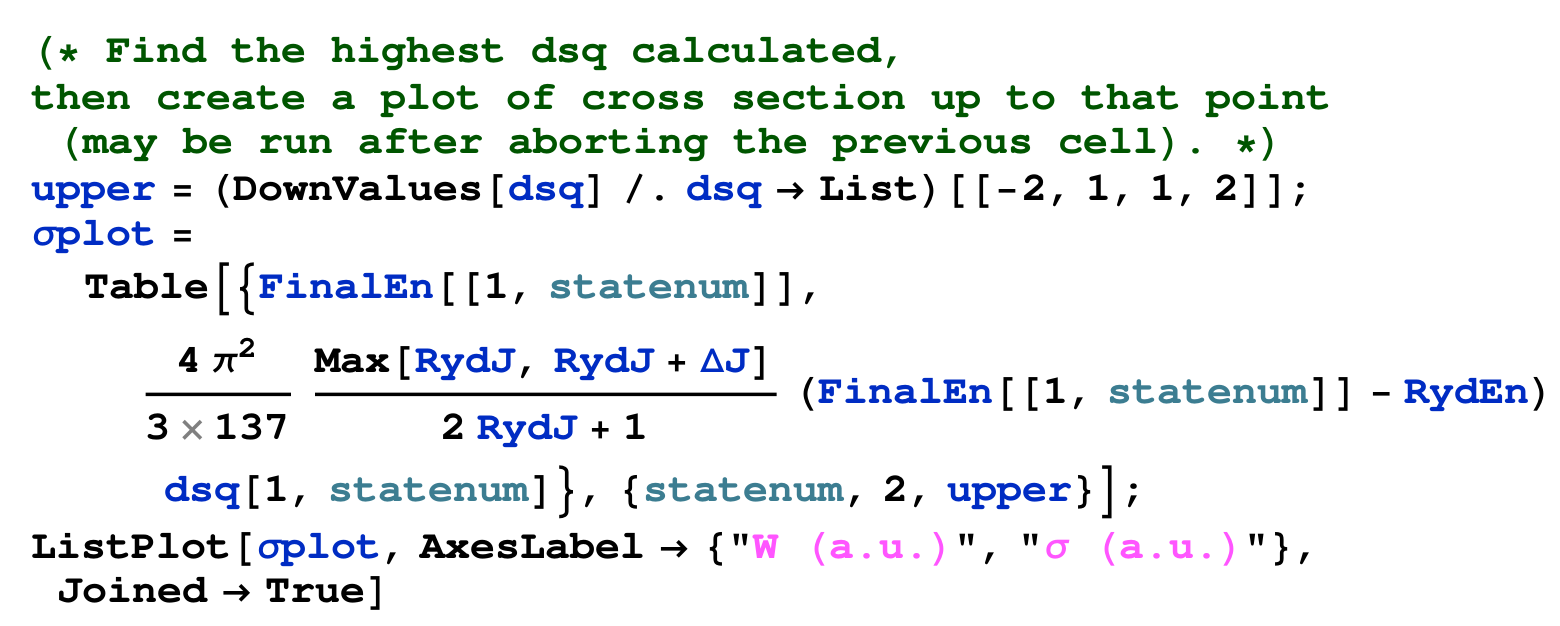}

\end{appendices}

\end{document}